\documentclass[sigconf]{acmart}
\usepackage{float}
\usepackage{siunitx}
\usepackage{subcaption}
\usepackage{enumitem}
\usepackage{subcaption}
\usepackage[utf8]{inputenc}
\usepackage{textalpha}
\usepackage{hyperref}

\usepackage{xcolor}
\usepackage[normalem]{ulem}
\newcommand{\new}[1]{\textcolor{black}{#1}}

\AtBeginDocument{%
  }

\copyrightyear{2026}
\acmYear{2026}
\setcopyright{cc}
\setcctype{by}
\acmConference[CHI '26]{Proceedings of the 2026 CHI Conference on Human Factors in Computing Systems}{April 13--17, 2026}{Barcelona, Spain}
\acmBooktitle{Proceedings of the 2026 CHI Conference on Human Factors in Computing Systems (CHI '26), April 13--17, 2026, Barcelona, Spain}
\acmPrice{}
\acmDOI{10.1145/3772318.3790389}
\acmISBN{979-8-4007-2278-3/2026/04}

\begin{document}

%%
%% The "title" command has an optional parameter,
%% allowing the author to define a "short title" to be used in page headers.
%%\title[CoLive]{CoLive: A VR-based Cognitive Empathy Training System using LLM-based Avatars and Dynamic Role-based Interactions}

%\title[CoEmpaTeam]{CoEmpaTeam: Training Cognitive Empathy using LLM-based Avatars and Dynamic Role Play in Virtual Reality}

\title[CoEmpaTeam]{CoEmpaTeam: Enhancing Cognitive Empathy using LLM-based Avatars and Dynamic Role Play in Virtual Reality}

%%
%% The "author" command and its associated commands are used to define
%% the authors and their affiliations.
%% Of note is the shared affiliation of the first two authors, and the
%% "authornote" and "authornotemark" commands
%% used to denote shared contribution to the research.
\author{Dehui Kong}
\email{dehui.kong@kit.edu}
\orcid{0009-0006-1703-1593}
\affiliation{%
  \institution{Karlsruhe Institute of Technology (KIT)}
  \city{Karlsruhe}
  \country{Germany}
}

\author{Martin Feick}
\email{martin.feick@kit.edu}
\orcid{0000-0001-5353-4290}
\affiliation{%
  \institution{Karlsruhe Institute of Technology (KIT)}
  \city{Karlsruhe}
  \country{Germany}
}

\author{Shi Liu}
\email{shi.liu@kit.edu}
\orcid{0000-0002-5714-3331}
\affiliation{%
  \institution{Karlsruhe Institute of Technology (KIT)}
  \city{Karlsruhe}
  \country{Germany}
}
  
\author{Alexander Maedche}
\email{alexander.maedche@kit.edu}
\orcid{0000-0001-6546-4816}
\affiliation{%
  \institution{Karlsruhe Institute of Technology (KIT)}
  \city{Karlsruhe}
  \country{Germany}
}

%%
%% By default, the full list of authors will be used in the page
%% headers. Often, this list is too long, and will overlap
%% other information printed in the page headers. This command allows
%% the author to define a more concise list
%% of authors' names for this purpose.
\renewcommand{\shortauthors}{Kong et al.}

%%
%% The abstract is a short summary of the work to be presented in the
%% article.
\begin{abstract}
Cognitive empathy, the ability to understand others‘ perspectives, is essential for effective communication, reducing biases, and constructive negotiation.
However, this skill is declining in a performance-driven society, which prioritizes efficiency over perspective-taking. Here, the training of cognitive empathy is challenging because it is a subtle, hard-to-perceive soft skill.
To address this, we developed \textit{CoEmpaTeam}, a VR-based system that enables users to train their cognitive empathy by using LLM-driven avatars with different personalities.
Through dynamic role play, users actively engage in perspective-taking, experiencing situations through another person's eyes.
\textit{CoEmpaTeam} deploys three avatars who significantly differ in their personality, validated by a technical evaluation and an online experiment (n=90). 
Next, we evaluated the system through a lab experiment with 32 participants who performed three sessions across two weeks, followed by a one-week diary study.
Our results showed a significant increase in cognitive empathy, which, according to participants, transferred into their real lives.
\end{abstract}

%%
%% The code below is generated by the tool at http://dl.acm.org/ccs.cfm.
%% Please copy and paste the code instead of the example below.
%%


%%
%% Keywords. The author(s) should pick words that accurately describe
%% the work being presented. Separate the keywords with commas.
%% A "teaser" image appears between the author and affiliation
%% information and the body of the document, and typically spans the
%% page.
\begin{teaserfigure}
  \includegraphics[width=\textwidth]{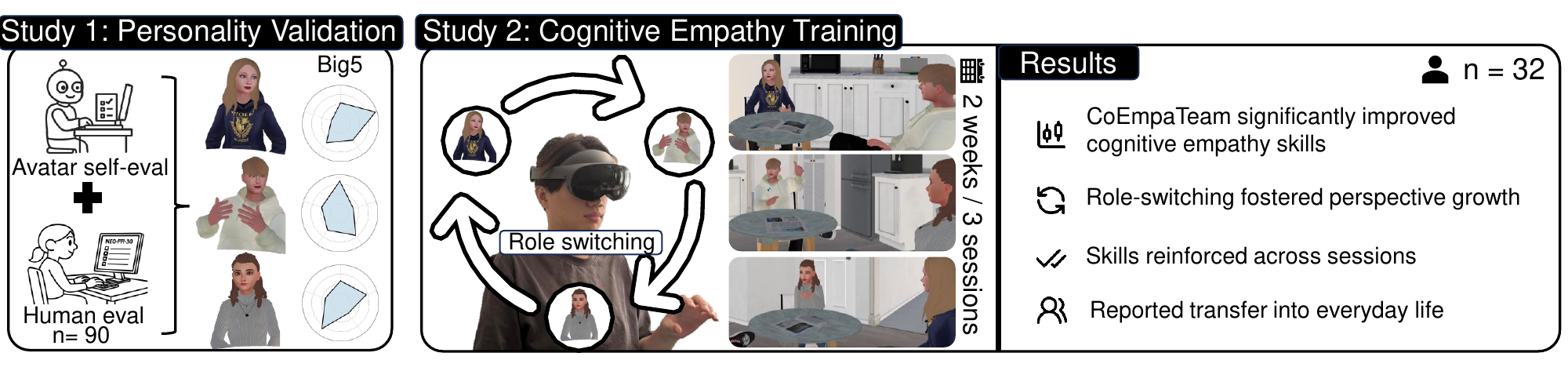}
    \caption{\emph{CoEmpaTeam} uses LLM-driven avatars with distinct personalities to foster cognitive empathy through role-switching. In Study 1, avatar self-assessment and human evaluation validated that Alice, Benji, and Caden reliably expressed their intended personalities (left). In Study 2, participants trained with \textit{CoEmpaTeam} across three sessions over two weeks, engaging in co-living task with different avatars (center). Results showed improvements in cognitive empathy and reported transfer of these skills into everyday life (right).}
    \Description{Overview of the CoEmpaTeam system and studies. Left: Study 1 validated avatar personalities through self- and human evaluation of Alice, Benji, and Caden. Center: Study 2 trained participants in role-switching with LLM-driven avatars in co-living tasks across three sessions. Right: Results show improved cognitive empathy, perspective growth, reinforcement across sessions, and transfer to everyday life.}
  \label{fig:teaser}
\end{teaserfigure}

%%\received{20 February 2007}
%%\received[revised]{12 March 2009}
%%\received[accepted]{5 June 2009}

%%
%% This command processes the author and affiliation and title
%% information and builds the first part of the formatted document.
\maketitle

\section{Introduction}

Cognitive empathy, often described as “knowing what another person is knowing,” highlights the importance of perspective taking~\cite{gencc2024situating}. It plays a crucial role in fostering effective communication, reducing biases, and enabling constructive negotiation outcomes~\cite{galinsky2005perspective}. Higher levels of empathy have also been associated with improved quality of life and well-being across diverse contexts~\cite{brazeau2010relationships}. However, research has shown that cognitive empathy among college students has declined by nearly 40\% over the past few decades~\cite{konrath2014changes}. While this decline is concerning, cognitive empathy is not a fixed trait but a malleable~\cite{gencc2024situating}, learnable~\cite{platt1994empathic}, trainable~\cite{lam2011empathy}, and transferable~\cite{van2018virtual} skill across different contexts. This combination of societal need and trainability underscores the importance of exploring new and more effective approaches to cultivating cognitive empathy.

Traditional approaches, such as classroom training programs~\cite{riess2012empathy, lam2011empathy} and online tutorials~\cite{sentas2018effects, kuhnley2023creatively}, have demonstrated only limited effectiveness, in part because they often lack relevance to learners’ real-life social contexts. Moreover, they face challenges of scalability: classroom training requires significant resources and logistical efforts (e.g., traveling to a training site), while online tutorials are typically video- or text-based, relying on passive learning and offering little embodied experience~\cite{land2004issues}. To address these limitations, emerging immersive technologies such as virtual reality (VR) offer unique affordances, including realistic immersion~\cite{herrera2018building}, customizable simulation~\cite{shin2018empathy}, and real-time interactivity~\cite{rovira2009use}. 
\new{These features allow users to experience another person’s perspective through role play. Compared with conventional in-person role play, VR enables controlled manipulation of perspectives~\cite{kishore2019virtual}, ensuring consistent scenario delivery across participants~\cite{herrera2018building} and scalable deployment without relying on multiple actors or participants~\cite{pan2018and}, which enables more situated engaging forms of cognitive empathy training~\cite{van2018virtual, bertrand2018learning, rifat2024cohabitant}.}

In this work, we adopt Davis’s definition of cognitive empathy: \textit{“the capacity to adopt another person’s perspective and understand their thoughts and feelings without necessarily sharing their emotional state”}~\cite{davis1983measuring}. This perspective-taking view has roots in Smith’s~\cite{smith2010theory} distinction between intellectual and emotional reactions to others’ experiences. While conventional role play supports perspective taking, it fixes participants into a single role, limiting opportunities to explore multiple viewpoints. To address this gap, we incorporate role-switching, a design mechanism that prompts participants to alternate between different roles, encouraging deeper reflection and a more nuanced understanding of others’ perspectives.

Cognitive empathy also emerges through rich social interaction rather than in isolation~\cite{gencc2024situating}. Effective empathy training, therefore, requires interactive partners that can respond dynamically and contextually to human behavior. The rapid advancement of large language models (LLMs) has made it possible to convincingly simulate diverse human behaviors and generate coherent, role-consistent responses~\cite{wan2024building, li2024exploring}. LLM-driven virtual avatars can provide natural language interaction and enhance social presence~\cite{christiansen2024exploring} through multimodal expressions such as gestures, facial expressions, and tone of voice~\cite{brito2025integrating}, offering an adaptive and immersive medium for cognitive empathy–focused interactions. To truly unlock VR’s potential as an~\textit{"empathy machine,"} such capabilities should be combined with targeted perspective-taking tasks and story prompts that actively engage participants in reflection and mentalizing~\cite{kukshinov2025seeing}.

\new{Given these gaps and opportunities, we frame our study around one overarching research question:}

\textbf{\new{RQ: How can we design a VR system that effectively supports the development of cognitive empathy?}}

\new{To address this overarching question, we examine three subquestions: (1) \textbf{RQ1}: how LLM-driven virtual avatars can be designed to exhibit distinct and coherent personalities; (2) \textbf{RQ2a}: how role-switching with LLM-driven avatars fosters cognitive empathy during VR training; and (3) \textbf{RQ2b}: how cognitive empathy skills developed during VR training translate into real-life social contexts.}

\new{To investigate these questions, we designed \textit{CoEmpaTeam}, a novel VR application that integrates role-switching with LLM-driven avatar interaction to create structured, multi-perspective social encounters and enhance cognitive empathy.} 
\textit{CoEmpaTeam} immerses users in a co-living scenario where they take on one of the three roles and, together with two avatars, engage in a retrospective meeting to revisit and renegotiate house rules. The task design was inspired by the Murder Mystery Roleplay, a form of live-action role play (LARP)~\cite{johansson2024larp}, and the avatar design was based on the Big Five personality framework~\cite{mccrae1992introduction} to capture a broad spectrum of everyday traits, with personalities validated through both LLM self-assessments and ratings from 90 participants in an online experiment. Using a mixed-methods approach and incorporating role-switching, we conducted three training sessions over the course of two weeks with 32 participants, followed by a one-week diary study. Our findings suggest that \textit{CoEmpaTeam} not only enhanced participants’ cognitive empathy during training but also facilitated the transfer of these skills into everyday life. In this work, we make \new{four} contributions:

\begin{itemize}
    \item We present \textit{CoEmpaTeam}, a VR-based system for training cognitive empathy, which enables users to engage in a co-living task with two LLM-driven avatars through a role-switching mechanism.  
    \item We design three avatar roles grounded in the Big Five personality theory and validate their personality consistency through both LLM-generated self-evaluations and assessments from 90 participants.
    \item We conduct a three-week study with 32 participants, including three training sessions over two weeks, followed by a one-week diary study, providing empirical evidence of the effectiveness and sustained impact of \textit{CoEmpaTeam}.
    \new{\item We release \textit{CoEmpaTeam} as an open-source system to support reproducibility, facilitate comparative evaluations, and enable extensions by the HCI research community.}

    % \item \textit{CoEmpaTeam}, a VR-based system for training cognitive empathy, enables users to engage in a co-living task with two LLM-driven avatars through role-switching.
    % \item A methodology for creating and validating theoretically grounded avatars, combining LLM-generated self-evaluations with human assessments.
    % \item Empirical evidence from a three-week study, including three training sessions over two weeks and a one-week diary, demonstrating the effectiveness and sustained impact of \textit{CoEmpaTeam}.
\end{itemize}

\section{Related Work}
\label{sec:RelatedWork}

\subsection{Approaches to Cultivating Cognitive Empathy in HCI}

Cognitive empathy concerns the ability to understand and recognize others’ emotions, circumstances, and experiences~\cite{boy2017showing, curran2019understanding}. In HCI, two approaches to fostering empathy are storytelling and role play~\cite{rifat2024cohabitant}. Storytelling is a method that “immerses the audience in the experiences of characters,” shaping memory and identity through shared narratives and thereby fostering empathetic understanding~\cite{mcadams2001psychology}. As a tool, storytelling has been applied across domains to enhance communication~\cite{hausknecht2019digitising}, build consensus and reconciliation~\cite{chongruksa2010storytelling}, and foster empathetic public engagement~\cite{loughran2022women}.

Beyond narrative immersion, role play provides an embodied pathway to empathy. By enacting different roles, participants can experience diverse perspectives~\cite{gieser2008embodiment} and enhance perspective-taking \cite{cooke2018empathic}. Research has shown that empathy is more likely to emerge when participants enact roles dissimilar to themselves~\cite{merilainen2012self}. However, the specific demands and situational constraints of different roles may also lead to divergent perspectives. Prior work suggests that exchanging social positions (i.e., role-switching) can help reconcile these divergences and deepen understanding~\cite{gillespie2011exchanging}. Moreover, role play generates episodic memories of lived experiences, which, when contrasted through role-switching, can further foster cognitive empathy~\cite{tulving1972episodic}. However, traditional role-play approaches typically center on a single role, restricting systematic perspective-switching across roles and limiting sustained opportunities for cognitive empathy training.

\subsection{Role of VR in Cognitive Empathy Training}

VR has been increasingly applied in HCI, demonstrating unique advantages in challenging contexts such as mental health~\cite{li2024exploring}, social inclusion~\cite{fernandez2025breaking}, and cross-cultural understanding~\cite{rifat2024cohabitant}. In recent years, VR has also been incorporated into empathy cultivation across diverse domains, including interfaith learning~\cite{rifat2024cohabitant}, crisis and illness experiences~\cite{kors2020curious,you2023bluevr}, human–animal relations~\cite{xu2024istraypaws}, cultural heritage~\cite{zhao2025immersive}, and intergenerational communication~\cite{shen2024legacysphere}. Despite the diversity of contexts, these applications emphasize immersive role play, contextualized interaction, and narrative guidance as core design elements to evoke cognitive engagement~\cite{muller2017through}.

VR has often been described as an \textit{“empathy machine”}~\cite{han2022immersive,hassan2020digitality}, yet its potential for cultivating empathy arises not from any single capability but from the integration of multiple affordances. First, realistic immersion enables users to inhabit specific perspectives and contexts~\cite{herrera2018building}, situating understanding within concrete social and cultural environments~\cite{rifat2024cohabitant}. Second, customized simulation allows designers to guide attention through spatial arrangements and narrative cues, helping participants engage with unfamiliar experiences and “step into others’ shoes”~\cite{shin2018empathy}. Finally, real-time interactivity provides a safe environment for practice and exploration~\cite{rovira2009use}. Compared to abstract representations, such embodied and concrete experiences are more likely to elicit critical reflection and empathic understanding, which can be deepened through repeated interaction~\cite{kolb2014experiential}.

Beyond these affordances, empirical studies provide evidence that VR role-switching can translate empathic experiences into real-world attitudes and behaviors~\cite{ma2020effects, van2018virtual}. For instance, role exchange in school bullying scenarios has been shown to foster moral reasoning and willingness to help~\cite{gu2022role}, while perspective shifts in police training enable trainees to move from the “perpetrator” role to the “victim” role, supporting experiential learning and empathy cultivation~\cite{kishore2019virtual}. These studies indicate that cognitive empathy is not only influenced by environmental context but can be intentionally cultivated through targeted task design. To fully realize VR’s potential, design must therefore go beyond immersion and narrative fidelity to integrate explicit cognitive interventions that scaffold perspective-taking and critical reflection~\cite{kukshinov2025seeing}. In this view, effective empathy training in VR is not about “watching a story” but about “inhabiting a role, driving the process, and making decisions.”

\subsection{LLM–Driven Social Avatars in VR}

Recent advances in LLMs have enabled their integration into VR, significantly enhancing interactive experiences in digital environments~\cite{aghel2024people, ozkaya2025llms}. 
Researchers have applied LLM-driven VR systems across domains such as education~\cite{liu2024classmeta,  cavallaro2024examining}, accessibility~\cite{lee2024gazepointar, cui2023glancewriter}, and training~\cite{li2025generative, fang2025social}.
For instance, \textit{ClassMeta} introduced adaptive classroom avatars that provide personalized learning support~\cite{liu2024classmeta}, while \textit{GlanceWriter} leveraged gaze-driven avatars to assist users with motor impairments in writing tasks~\cite{cui2023glancewriter}. In training contexts, generative social simulations have been used to create realistic social scenarios that support self-care practice under stress~\cite{fang2025social}. 
Across many of these domains, LLM-driven avatars have emerged as the central medium for interaction, capable of generating context-relevant dialogues by integrating system prompts, conversational memory, and user input~\cite{llanes2024developing}.
These developments build on a longer trajectory of virtual agent research. 
Early systems evolved from scripted characters~\cite{kacmarcik2006using} to emotionally driven agents such as \textit{FearNot!}, which generated emergent narratives to foster empathy in anti-bullying education~\cite{aylett2005fearnot}.
More recently, generative approaches such as \textit{Generative Agents} have demonstrated how LLM-powered characters can sustain daily routines and interpersonal relationships in sandbox environments~\cite{park2023generative}. 
This trajectory from early scripted logic to contemporary LLM-powered avatars reflects a broader shift toward open-ended, generative modes of interaction, while recent work has also begun extending these capabilities into more structured training scenarios.

Building on this trajectory, LLM-driven avatars can dynamically adjust narratives, emotions, and motivations, thereby delivering more immersive and authentic interactive experiences~\cite{li2024exploring,zhu2025designing}. In social VR, they enhance presence and social realism~\cite{guimaraes2020impact,shoa2023sushi}; in medical and training contexts, they foster engagement through emotionally rich feedback~\cite{zhu2025designing}. Beyond language generation, their credibility also depends on multimodal cue integration, visual appearance, and personality design~\cite{cassell2001embodied,van1998persona}. Building on these design considerations, recent studies have explored applications ranging from virtual patient simulators~\cite{zhu2025designing} to virtual classroom companions~\cite{liu2024classmeta} and language-learning assistants~\cite{pan2025ellma}, demonstrating the potential of LLM-driven avatars across diverse contexts.

In terms of interaction forms, LLM-driven avatars have expanded from text to multimodal expressions such as gestures, facial expressions, and gaze, enabling more immersive and credible interactions~\cite{normoyle2024using,qing2023story}.
This shift is pushing VR avatars beyond scripted dialogues toward dynamic, context-aware, and emotionally resonant modes of interaction.
A central mechanism is the temporal and spatial alignment of verbal and visual cues.
Such alignment supports conversational fluency and spatial orientation, reduces cognitive load through sensory consistency, and enhances task performance~\cite{cioffi2025speech,martin2022multimodality}.
In VR environments, these multimodal interactions not only improve intuitiveness but also foster stronger emotional and social awareness, thereby deepening user engagement~\cite{guimaraes2020impact, han2022immersive}. However, despite these advances, existing LLM-driven avatar systems rarely adopt distinct personalities across multiple avatars, limiting their capacity to present differentiated perspectives.

Overall, prior work highlights the potential of role play, VR, and LLM-driven avatars for empathy training, yet these elements have rarely been integrated into a unified and systematic framework. Building on these insights, we introduce \textit{CoEmpaTeam}, a VR system that unites multi-avatar interaction with dynamic role-switching to advance situated training for cognitive empathy.
\section{Developing \textit{CoEmpaTeam}}

\textit{CoEmpaTeam} is an immersive VR application in which participants engage in role play with two LLM-driven avatars, each embodying distinct personalities within a designed living situation. The scenarios situate participants in everyday, conflict-prone contexts that require communication, negotiation, perspective-taking, and joint decision-making. By switching roles among these characters, participants directly experience multiple perspectives and practice reconciling differences through dialogue and negotiation, which are essential for understanding, anticipating, and responding to others’ perspectives in real-world social contexts. Through this process, \textit{CoEmpaTeam} aims to cultivate users' cognitive empathy.

\subsection{Design}

\subsubsection{Co-living Task}

We designed \textit{CoEmpaTeam} to immerse participants in a familiar yet conflict-prone context—shared living arrangements~\cite{clark2020managing, foulkes2019impact}. This choice is motivated by the fact that co-living often involves disagreements over noise, cleanliness, kitchen use, guest policies, and personal boundaries~\cite{kim2023design, clark2020managing, foulkes2019impact}. Such everyday conflicts are highly relatable to daily experiences while also laden with tensions around value differences and responsibility allocation. Inspired by role-playing games such as Murder Mystery Roleplay~\cite{johansson2024larp, bowman2014educational}, we translated this scenario into a co-living task, where participants and two avatars hold a retrospective house meeting to decide whether to continue co-living and how to renegotiate house rules (see Supplement). 

In this process, participants must negotiate from multiple standpoints, actively engaging in perspective-taking. Psychological research regards perspective-taking as a critical mechanism for cultivating cognitive empathy~\cite{davis1983measuring}, as it enhances self–other overlap and facilitates social coordination~\cite{galinsky2005perspective}. Building on this foundation, we hypothesize that the everyday conflicts embedded in \textit{CoEmpaTeam} provide an effective basis for eliciting perspective-taking, thereby fostering the development of cognitive empathy.

\subsubsection{Role Play}

\textit{CoEmpaTeam} extends conventional role play with a role-switching mechanism that requires participants to sequentially enact three different characters across multiple sessions. Before each session, participants receive detailed role cards containing Basic Info, Lifestyle Log, Hidden Motivation, and Stance on House Rules (see Supplement), which provide sufficient context for role immersion.
By alternating between roles, participants move beyond a single viewpoint and are encouraged to confront conflicting perspectives. This design transforms role play into a structured multi-perspective exercise, prompting reflection, re-examination of assumptions, and deeper understanding of others’ positions, thus operationalizing prior findings on perspective-taking as a concrete mechanism for cultivating cognitive empathy~\cite{galinsky2005perspective}.

\subsubsection{Avatar Design}
\label{sec: Avatar Design}

Personality as a driven variable was central to the design of the avatars in the \textit{CoEmpaTeam}. The Big Five personality framework~\cite{goldberg2013alternative}, widely validated in psychology and increasingly adopted in HCI to capture individual differences~\cite{buck2023avatar, ju2025toward}, served as the basis for our design. Each avatar was modeled with a distinct personality profile based on the Big Five~\cite{goldberg2013alternative}, varying in openness, conscientiousness, extraversion, agreeableness, and neuroticism (see Supplement). This variation was designed to foster diverse perspectives, communication styles, and interpersonal tensions, conditions that are critical for eliciting cognitive empathy and perspective-taking, which form the training goals of the task. To operationalize these traits, we employed LLM prompting strategies. Following a zero-shot learning approach~\cite{brito2025integrating}, we crafted instructions that implicitly conveyed the desired traits, guiding the model to generate responses aligned with specific personality characteristics. Prior work demonstrates that descriptive prompts can elicit stable expressions of personality by framing the interaction context in ways that encourage the model to naturally embody the intended persona~\cite{jiang2023personallm}. Our implementation builds on these insights to ensure that avatar dialogue consistently reflects distinguishable yet coherent personalities.

To situate these personalities within a manageable yet socially rich context, the avatars were arranged in a three-person group. Triads represent one of the fundamental “core configurations” of human social interaction~\cite{caporael1997evolution} and strike an effective balance between complexity and cognitive load~\cite{sweller2011cognitive}, providing enough diversity for perspective-taking while avoiding cognitive overload. For narrative coherence, avatars were assigned fixed genders (two females, one male). While gender was not treated as a study variable, it remains unclear whether this choice influenced participants’ perceptions. Future work should examine how different gender configurations may shape experiences in similar settings. 

To further enhance immersion, we integrated verbal and non-verbal cues into avatar behavior. Prior work highlights that the combination of speech and embodied signals is essential for creating natural and engaging human–agent interactions~\cite{cassell2001embodied}. Building on this, the LLM output was structured into four fields: speaker, text, emotion, and gesture. These were mapped to avatar behaviors (see \ref{sec: Implementation}), ensuring that linguistic content was consistently accompanied by expressive cues and that verbal output aligned with corresponding emotional and gestural expressions. To avoid avatars appearing static or mechanical, we further introduced micro-behaviors such as random blinking and subtle eye movements, thereby enhancing social presence and strengthening both clarity and \new{expressiveness} in interaction~\cite{wu2021using}.

\begin{figure}[ht]
  \centering
  \includegraphics[width=0.7\linewidth]{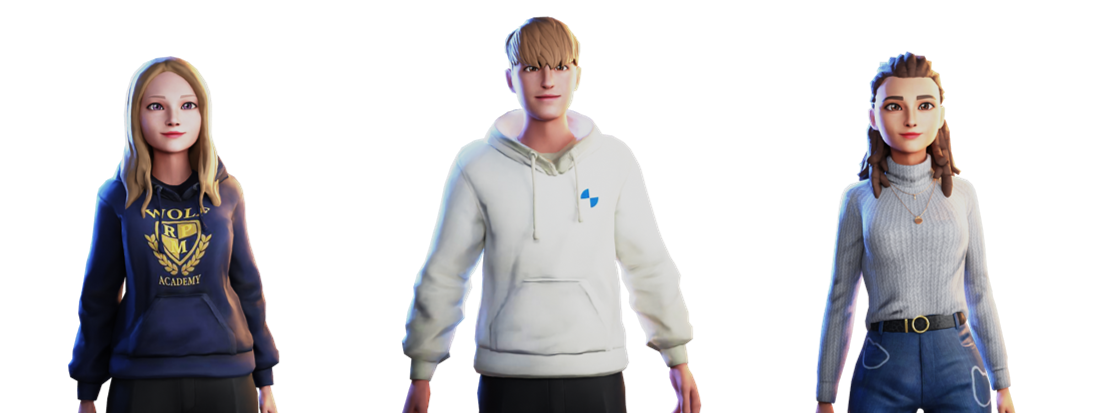}
  \caption{Avatars used in the \textit{CoEmpaTeam} system. From left to right: Alice, Benji, and Caden.}
  \Description{Three avatars used in the \textit{CoEmpaTeam} system, shown from left to right: Alice, Benji, and Caden, each with distinct visual appearance representing different personalities.}
  \label{fig: avatar}
\end{figure}

\subsubsection{Implementation}
\label{sec: Implementation}
\textit{CoEmpaTeam} was implemented in Unity3D (2022.\allowbreak 3.\allowbreak 51f1)\footnote{https://unity.com/} and deployed on the Meta Quest Pro headset\footnote{https://www.meta.com/quest/quest-pro/}. The virtual environment was set in a kitchen to reflect the domestic context of co-living. Most 3D assets were sourced online and optimized to balance authenticity with the limited computational resources of standalone VR devices. Avatar models were generated using Ready Player Me\footnote{https://hub.readyplayer.me/avatar/choose} (see \autoref{fig: avatar}).

During interaction, participants’ speech was transcribed into text using a locally deployed OpenAI Whisper model\footnote{https://github.com/openai/whisper}. Transcriptions were appended to the dialogue history and forwarded to a Python backend, where structured prompts were constructed. These prompts incorporated task background, turn-taking rules, output format, and avatar-specific personas, including their possible gestures and emotional repertoires (see Supplement). The backend then queried the Llama 3.1 8B Instruct model\footnote{https://huggingface.co/meta-llama/Llama-3.1-8B-Instruct} to generate a response. The structured output was sent to the Unity client, where it was parsed to drive avatar behaviors. Each response was structured into four fields: \textit{speaker}, \textit{text}, \textit{gesture}, and \textit{emotion}.
The \textit{text} field was converted into natural speech via ElevenLabs TTS\footnote{https://elevenlabs.io/app/developers/api-keys}, with articulation synchronized using the Oculus LipSync Unity SDK\footnote{https://developers.meta.com/horizon/documentation/unity/audio-ovrlipsync-unity/}. \textit{Gestures} were animated with Mixamo\footnote{https://www.mixamo.com/}, while \textit{emotions} were mapped to avatar-specific facial blendshapes to reflect distinct personalities. To further enhance social presence, avatars were equipped with dynamic gaze control using Unity Final IK (Look At IK)\footnote{https://docs.readyplayer.me/ready-player-me/integration-guides/unity/setup-for-xr-beta/setup-final-ik}, enabling them to orient toward the current speaker. Code and the Unity package are available online\footnote{\url{https://github.com/kindhui62/CoEmpaTeam}}.

\subsubsection{\textit{CoEmpaTeam} Role-play Workflow}
\begin{figure}[ht]
  \centering
  \includegraphics[width=1\linewidth]{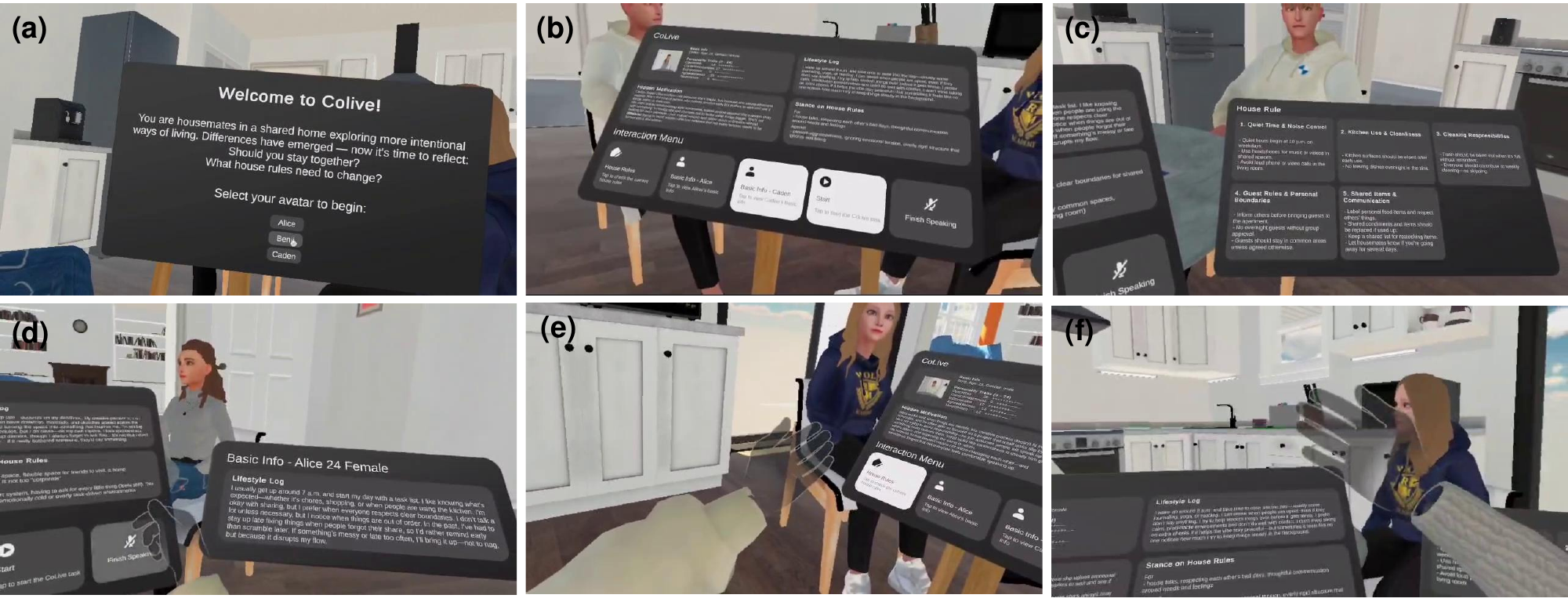}
  \caption{\textit{CoEmpaTeam} system interface and role-play workflow: (a) Welcome screen for avatar selection. (b) Role-specific information with options to start the task or finish speaking. (c) Shared house rules organized into five categories. (d) Example of basic information for Alice to support role-play. (e-f) Motion-synchronized avatar arms when embodying Benji and Caden (Alice analogous).}
  \Description{Screenshots of the CoEmpaTeam system interface and role-play workflow. Panel (a) shows the welcome screen with avatar selection. Panel (b) displays role-specific information and task controls. Panel (c) presents shared house rules organized into categories. Panel (d) illustrates role background information for Alice. Panels (e–f) show motion-synchronized avatar embodiment when playing Benji and Caden.}
  \label{fig: systemandworkflow}
\end{figure}

Upon entering the environment, participants selected a role (see \autoref{fig: systemandworkflow}a) and reviewed information about their assigned character, including Big Five personality traits, lifestyle logs, and hidden motivations (see \autoref{fig: systemandworkflow}b). To support role-play, they were also given basic information about the two other avatars (see \autoref{fig: systemandworkflow}d). Embodiment was reinforced through motion-synchronized avatar arms (see \autoref{fig: systemandworkflow}e-f), enhancing visuomotor alignment and presence~\cite{tham2018understanding}. This design was informed by the Proteus effect, which suggests that embodying a virtual character can shape users’ attitudes and behaviors~\cite{yee2007proteus}.

To initiate the task, participants clicked “start.” After each speech turn, they pressed “finish speaking” to transmit their input to the system (see \autoref{fig: systemandworkflow}b), which in turn triggered responses from the two avatars. This interaction loop structured the role-play while maintaining flexibility, ensuring consistent turn-taking and supporting naturalistic conversations with the avatars.

%%
%%\begin{figure}[ht]
%%  \centering
%%  \includegraphics[width=1\linewidth]{Figure/interaction.pdf}
%%  \caption{\textit{CoEmpaTeam} system interface and role-play workflow: (a) Welcome screen for avatar selection. (b) Role-specific information with options to start the task or finish speaking. (c) Shared house rules organized into five categories. (d–e) Basic information of the other two avatars to support role-play. (f–h) Motion-synchronized avatar arms when embodying Alice, Benji, and Caden.}
%%  \Description{CoEmpaTeam system interface and role-play workflow. (a) Welcome screen for avatar selection. (b) Role-specific information with options to start the task or finish speaking. (c) Shared house rules organized into five categories. (d–e) Basic information of the other two avatars to support role-play. (f–h) Motion-synchronized avatar arms when embodying Alice, Benji, and Caden.}
%%  \label{fig: systemandworkflow}
%%\end{figure}

\subsection{Study 1: Validation of Avatars}

\new{To address RQ1, we combined avatar self-assessment with human evaluation to validate whether the three designed avatars exhibited behaviors consistent with their intended personalities.}

\subsubsection{Avatar Self-assessment} Following~\cite{kroczek2025influence}, we administered the NEO-FFI-30 personality inventory~\cite{korner2008personality} to each avatar, repeating the procedure 100 times to account for variability in model outputs. We calculated mean scores ($M$) and standard deviations ($SD$) for the Big Five traits, as well as cosine similarity to the intended profiles as a measure of overall fidelity (as detailed in \autoref{tab: avatarself}). Alice’s responses were highly consistent with her predefined profile (cosine similarity $M=0.997$, $SD=0.002$), reflecting her structured and rule-oriented persona. Benji’s profiles, in contrast, showed greater variability ($M=0.862$, $SD=0.017$), aligning with his more flexible and less organized character. Caden fell between these extremes, exhibiting high fidelity ($M=0.950$, $SD=0.010$) while maintaining slight natural variation. Together, these results demonstrate that the avatars not only reproduced their intended Big Five configurations but also expressed variability patterns consistent with their designed personalities.

\subsubsection{Human Evaluation} To evaluate whether the avatars’ overall behaviors aligned with their intended personalities, we conducted a between-subjects online study. 90 participants were randomly assigned to one of three conditions (focusing on either Alice, Benji, or Caden), with 30 participants evaluating each avatar. Each participant watched a five-minute pre-recorded video of a discussion round between the avatars before completing a third-person version of the NEO-FFI-30 personality inventory~\cite{korner2008personality} and an additional five-item Likert questionnaire (1–5) assessing the perceived consistency between the avatar’s verbal expressions, non-verbal behaviors, and vocal style (see Supplement).

\begin{table}[t]
\centering
\caption{Big Five trait profiles of avatars across 100 NEO-FFI-30 trials ($M \pm SD$), with cosine similarity to target profiles.}
\label{tab: avatarself}
\resizebox{\linewidth}{!}{
    \begin{tabular}{lcccccc}
    \toprule
     & Openness & Conscientiousness & Extraversion & Agreeableness & Neuroticism & Cosine Similarity \\
    \midrule
    Alice (Target) & 13 & 23 & 13 & 15 & 6 & -- \\
    Alice ($M \pm SD$) & $13.83 \pm 1.38$ & $22.95 \pm 0.28$ & $14.83 \pm 0.64$ & $14.39 \pm 1.22$ & $6.16 \pm 0.83$ & $0.997 \pm 0.002$ \\
    \midrule
    Benji (Target) & 20 & 9 & 17 & 13 & 12 & -- \\
    Benji ($M \pm SD$) & $19.79 \pm 1.12$ & $8.94 \pm 1.17$ & $16.81 \pm 1.14$ & $13.04 \pm 1.09$ & $12.03 \pm 1.06$ & $0.862 \pm 0.017$ \\
    \midrule
    Caden (Target) & 14 & 17 & 8 & 20 & 10 & -- \\
    Caden ($M \pm SD$) & $13.79 \pm 1.12$ & $16.88 \pm 0.67$ & $7.81 \pm 1.14$ & $20.04 \pm 0.67$ & $10.03 \pm 1.06$ & $0.950 \pm 0.010$ \\
    \bottomrule
    \end{tabular}
}
\end{table}

We recruited 90 participants (42 male, 48 female) through Prolific\footnote{\url{https://www.prolific.com}}, an online participant recruitment platform. Each participant received €3 as compensation. To avoid potential transfer effects from prior exposure to the avatars, none of the participants in this validation study took part in the subsequent main experiment. On average, the study took 12 minutes to complete.

To ensure data quality and align with established gold-standards for crowdsourced research~\cite{gadiraju2015understanding, kittur2008crowdsourcing}, the questionnaire included a attention-check item (e.g., “Please select ‘strongly agree’ for this statement”), which all retained submissions passed. We also examined completion times to identify potential low-effort responses, such as unusually short durations; no such anomalies were detected.

\begin{figure}[ht]
  \centering
  \includegraphics[width=1\linewidth]{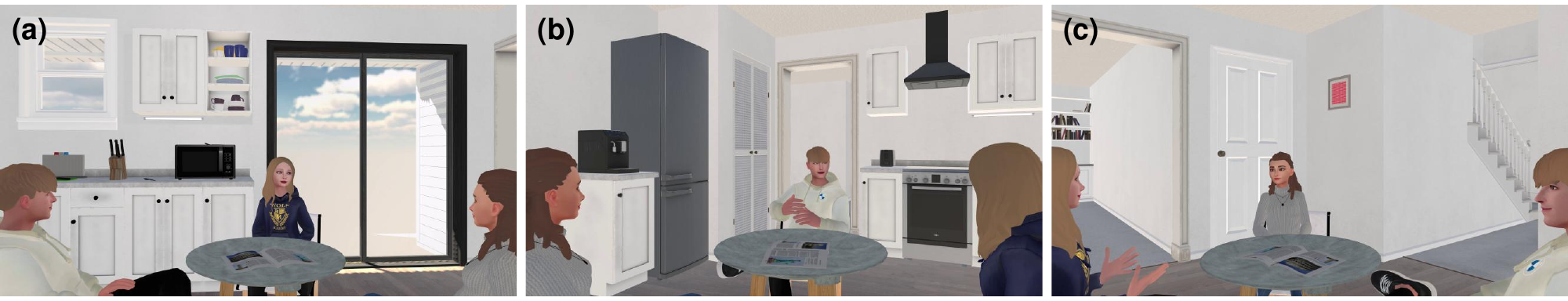}
  \caption{Example frames from the evaluation videos. Each condition featured a different target avatar placed in the center of the scene: (a) Alice, (b) Benji, and (c) Caden. The other two avatars were positioned to the sides to simulate a realistic group interaction while directing participants’ attention to the evaluated role.}
  \Description{Example frames from evaluation videos showing different target avatars in group interaction scenes. Panel (a) centers on Alice, panel (b) on Benji, and panel (c) on Caden, with the remaining avatars positioned to the sides to create a realistic group setting.}
  \label{fig:onlinestudyvideo}
\end{figure}

Participants first watched a five-minute pre-recorded video in which the three avatars discussed the co-living task. In each video, the target avatar was placed in the center of the screen, with the other two positioned to the sides to simulate a realistic group interaction while directing participants’ attention to the evaluated character (see \autoref{fig:onlinestudyvideo}). After viewing, participants completed the questionnaires described above.

\subsubsection{Results} 

We collected 90 complete questionnaire responses. For each avatar, we aggregated personality ratings and compared them with the predefined trait profiles. To evaluate reliability, we assessed the third-person version of the NEO-FFI-30~\cite{korner2008personality}. Across avatars, the subscales showed acceptable to good internal consistency (Cronbach’s α = .67–.87), with most exceeding the conventional .70 threshold. The only exception was Openness for Alice (α = .67), which remains acceptable for short scales such as the NEO-FFI-30~\cite{schmitt1996uses}.  

We then analyzed the alignment between perceived and predefined personalities using two complementary measures: 
(1) Pearson’s $r$, capturing whether the relative ranking of trait scores across the Big Five was consistent, and 
(2) Cosine similarity, assessing the overall configurational similarity between the perceived and predefined profiles as vectors. 
Pearson’s $r$ thus captures trait ranking alignment, while cosine similarity reflects overall configurational fidelity.
Applying these measures, we found that participants’ perception of Alice was strongly and significantly correlated with the intended profile ($r = .93$, $p < .05$). Benji’s ratings also indicated a strong correlation ($r = .87$, $p = .052$). Similarly, Caden’s perceived profile demonstrated a strong and significant correlation ($r = .89$, $p < .05$). Cosine similarity analyses confirmed the overall alignment: Alice ($M = 0.969$, $SD = 0.027$), Benji ($M = 0.959$, $SD = 0.025$), and Caden ($M = 0.967$, $SD = 0.036$).  Finally, we calculated mean (M) and standard deviation (SD) scores for each item of the verbal–nonverbal and vocal consistency questionnaire. Ratings were positive overall, with mean scores ranging between 3.5 and 4.1 (out of 5) (see Supplement). Participants rated the avatars highest on items concerning the match between voice and personality as well as the overall believability of the character (Q3–Q4). Overall, these findings indicate that participants perceived all three avatars as internally consistent and believable, providing validation of the character designs.

\subsubsection{Summary}

Our dual evaluation, combined with the avatar self-assessment and human evaluation, confirms that the avatars consistently expressed their intended personalities, Alice as structured and rule-oriented, Benji as flexible and less organized, and Caden as accommodating and harmony-seeking. This dual validation establishes the credibility of the avatar designs, ensuring the avatars are both computationally coherent and experientially believable. This provides a solid foundation for the subsequent cognitive empathy training study.
\section{Study 2: Cognitive Empathy Training}

\new{To address RQ2a and RQ2b, we conducted a three-week study in which participants completed three role-switching sessions across two weeks (RQ2a), followed by a one-week diary study (RQ2b), to examine how repeated role-switching with \textit{CoEmpaTeam} fosters cognitive empathy and how these skills transfer into everyday life.}

\subsection{Study Design}

Each participant engaged in three training sessions, role-playing Alice, Benji, and Caden. To account for the distinct traits of the roles while minimizing potential order effects, we employed a 3×3 Latin square design that counterbalanced the sequence in which participants enacted the roles.
Following prior VR empathy research in HCI~\cite{kors2020curious, rifat2024cohabitant}, which evaluates systems holistically, we likewise examine \textit{CoEmpaTeam} as an integrated training experience rather than isolating individual components.

The training phase lasted two weeks, with sessions scheduled at least two days apart to encourage reflection while maintaining continuity. At the beginning of each session, participants were screened for VR-related motion sickness and mental health issues to ensure safe participation. Completion of all three sessions was required for inclusion in the study. Before and after the training sessions, participants completed questionnaires to assess cognitive empathy and related constructs (see \ref{sec: datacollection}). 
After the last training session, we invited a subset of participants for a 20-minute semi-structured interview about their experiences with \textit{CoEmpaTeam}. 

To measure the real-world transfer of cognitive empathy training, participants completed a three-entry diary study (one entry every two days) during the week after the training. Diary methods are particularly valuable in HCI research for their ecological validity, as they enable in-situ reflection on real-world experiences beyond the controlled lab environment~\cite{czerwinski2004diary}.

\subsection{Participants}

We recruited 32 participants (22 male, 10 female) from a local university, aged 18–34 ($M = 24.00$, $SD = 3.41$). All attended the first session, 27 returned for the second, and 23 completed the third. The final analysis included 22 participants (14 male, 8 female, $M = 24.39$, $SD = 3.71$) who completed all three training sessions. One additional participant was excluded due to exceeding the cut-off on the General Health Questionnaire–12 (GHQ-12)~\cite{del200812}, consistent with our pre-defined screening criteria. Based on participants’ feedback, attrition appeared to be mainly due to scheduling conflicts (e.g., overlapping classes or exams), rather than study-related discomfort, though other contributing factors cannot be fully ruled out. 
The cultural backgrounds of participants were diverse, with 9 participants from Western Europe, 3 from Eastern Europe, 4 from East Asia, 5 from South Asia, and 1 from the Middle East.
Regarding prior VR experience, 2 participants had never used VR, 14 had used it once or twice, and 6 reported occasional use a few times per year. None reported frequent VR use.

Participants received increasing compensation across sessions (€13, €15, €17) to acknowledge their time commitment, with additional payments for the interview (€5) and for each of the three diary entries (€3, €4, €5). The study was approved by the institutional ethics and data protection committee, and all procedures adhered to applicable data protection regulations. Participants were informed that participation was voluntary across all stages of the study and that they could withdraw at any time without penalty.

\subsection{Procedure}

\subsubsection{Training Sessions}

At the beginning of the study, each participant was assigned to an individual room to ensure a quiet environment. After providing informed consent, participants completed a 15-minute onboarding session to familiarize themselves with the system. They then received printed role materials describing the co-living task, including background information, task instructions, house rules and a role card outlining personality traits, lifestyle log, hidden motivations, and stance on house rules (see Supplement). To support role immersion, we also provided an overview of the Big Five personality dimensions (see Supplement). Participants subsequently engaged in a short practice phase to get familiar with the VR headset and system.

Each participant completed three training sessions across two weeks. At the beginning and end of each session, participants filled out pre- and post-test questionnaires (see \ref{sec: pretest}, \ref{sec: posttest}). During the session, they wore a Meta Quest Pro headset, selected their assigned role, and engaged in a 20-minute discussion on household rules with two other LLM-driven avatars, each with distinct personalities. The order of roles (Alice, Benji, and Caden) was counterbalanced across participants.

\subsubsection{Post-Training Interview \& Diary}

After completing all three sessions, participants were invited to a voluntary 20-minute semi-structured interview focusing on their overall experience with the \textit{CoEmpaTeam} system, engagement in cognitive empathy-related practices, and perceptions of the role-play and avatars. In the following week, they participated in a diary study consisting of three online entries (one every two days), guided by open-ended prompts. The diaries encouraged participants to reflect on empathy-related experiences in their daily life and to consider possible links to the VR sessions (see \ref{sec: qualitative}).

\subsection{Data Collection}
\label{sec: datacollection}

\subsubsection{Pre-test Questionnaire}
\label{sec: pretest}

At the beginning of the study, participants provided demographic information (age, gender, nationality, and prior VR experience). Before each training session, we administered the General Health Questionnaire (GHQ-12)~\cite{del200812} to screen for mental health risks and the Simulator Sickness Questionnaire (SSQ)~\cite{kennedy1993simulator} to monitor VR-related motion sickness. These measures were used solely for screening and did not enter the analysis.

To assess cognitive empathy, we used the Interpersonal Reactivity Index (IRI)~\cite{davis1980multidimensional}, a widely used self-report measure of empathy. The IRI consists of four subscales: Perspective Taking, Empathic Concern, Fantasy, and Personal Distress. Consistent with our research focus, we examined the two subscales most directly linked to cognitive empathy: Perspective Taking (PT) and Fantasy (FS). PT captures the tendency to adopt others’ viewpoints in everyday life, whereas FS reflects the tendency to imaginatively identify with fictional characters, paralleling the role-taking required in our VR training. Both subscales have demonstrated high internal consistency in prior research (PT: Cronbach’s $\alpha = .83$; FS: $\alpha = .86$)~\cite{kelly2022borderline}.

Participants responded to 14 items on a 5-point Likert scale (1 = does not describe me well, 5 = describes me very well), with higher scores indicating stronger cognitive empathy. Example items include \textit{“I try to look at everybody’s side of a disagreement before I make a decision”} (PT) and \textit{“I often find myself imagining what it would be like to be in the shoes of a character in a book/movie.”} (FS). The IRI has been widely used to detect changes in empathy using paired t-tests~\cite{davis1983measuring, wu2023efficacy} and has also been applied in VR research to investigate cognitive empathy and perspective-taking~\cite{herrera2018building, van2018virtual}, making it a well-established and reliable measure for our study.

\subsubsection{Post-test Questionnaire}
\label{sec: posttest}

After each training session, participants completed several measures. To assess changes in cognitive empathy, we administered the IRI~\cite{davis1980multidimensional} again. To evaluate experiences in the virtual environment, we included the Presence Questionnaire (PQ, 5-point Likert scale)~\cite{witmer1998measuring} and the ownership subscale of the Virtual Embodiment Questionnaire (VEQ, 5-point Likert scale)~\cite{gonzalez2018avatar}. To explore whether the avatars elicited an uncanny valley effect, we added the Uncanny Valley Index (UVI, 7-point Likert scale)~\cite{ho2010revisiting}. Finally, after the third session, participants completed the NASA Task Load Index (NASA-TLX, 7-point Likert scales)~\cite{hart1988development} to examine potential task-related strain.

\subsubsection{Qualitative Data}
\label{sec: qualitative}

To complement the quantitative measures, we collected qualitative data through semi-structured interviews and diary entries to capture participants’ perspectives in depth.
The post-training interviews explored participants’ overall experience with the system, their engagement in cognitive empathy (e.g., perspective-taking and role-switching), their sense of role immersion, perceptions of the avatars, and reflections on possible improvements to the empathy training mechanisms. 
In the following week, participants completed a diary study consisting of three entries (one every two days). Each entry included four open-ended prompts and followed a shared structure: participants were asked to recall a recent real-world interpersonal interaction, describe how they engaged in perspective-taking, and reflect on whether any aspects of the VR training informed their experience. 
While the overall structure was consistent, each entry varied slightly in reflective emphasis. The first entry focused on recalling a recent interaction and describing how they attempted to understand another person’s viewpoint. The second centered on how participants adapted their responses during interactions and whether role-switching in VR influenced this process. The third invited broader reflection on changes over time, including perceived changes in cognitive empathy related behaviors and intentions to continue applying related strategies. The full set of diary prompts is included in the Supplement.
To ensure data quality, participants were reminded that there were no right or wrong answers, that compensation was based solely on completion rather than content. 
Together, the interviews and diaries captured participants’ reflections on cognitive empathy-related experiences inside and outside VR.
We analyzed all qualitative data using thematic analysis~\cite{braun2006using}, following an inductive coding approach~\cite{ryan2003techniques} to identify recurring themes.

\subsection{Results}

We report both quantitative and qualitative findings. Quantitatively, we focus on changes in cognitive empathy using the Interpersonal Reactivity Index (IRI), alongside measures of presence (PQ), embodiment (VEQ), uncanny valley perceptions (UVI), and task load (NASA-TLX). Qualitatively, our analysis of the interviews and diary entries provides deeper insights into participants’ training experiences and the transfer of empathy strategies into daily life.

\subsubsection{Quantitative Results}
\label{sec: quantitativeresults}

\paragraph{Cognitive Empathy: Perspective Taking and Fantasy (IRI)}

\begin{figure}[ht]
  \centering
  \begin{subfigure}[t]{0.45\linewidth}
    \centering
    \includegraphics[width=\linewidth]{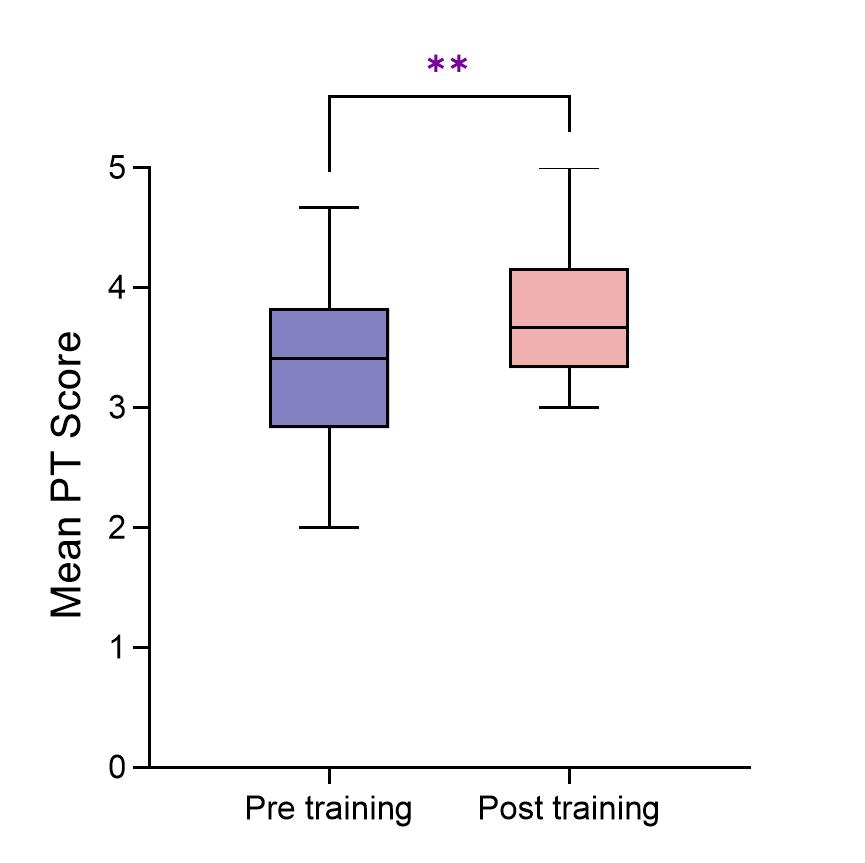}
    \caption{PT}
  \end{subfigure}
  \begin{subfigure}[t]{0.45\linewidth}
    \centering
    \includegraphics[width=\linewidth]{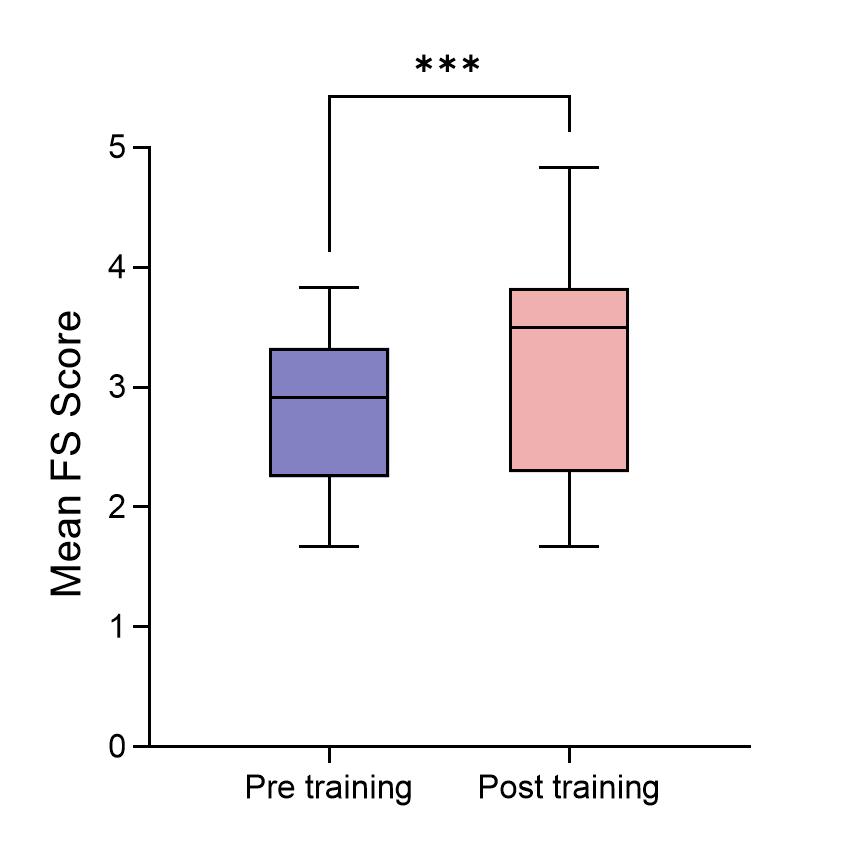}
    \caption{FS}
  \end{subfigure}
  \caption{Pre- and post-training changes in IRI subscales. Both PT and FS increased significantly.}
  \Description{Boxplots of pre- and post-training scores for Interpersonal Reactivity Index subscales. Panel (a) shows Perspective Taking, and panel (b) shows Fantasy. Both measures increased significantly after training, indicating improved cognitive empathy.}
  \label{fig: ptfs}
\end{figure}

The IRI sample included 22 participants. In line with our focus on cognitive empathy, we analyzed the Perspective Taking (PT) and Fantasy (FS) subscales separately. Both subscales demonstrated high internal consistency in our sample (Cronbach’s $\alpha = .77$–$.86$; FS: $\alpha = .79$–$.87$). Paired-sample t-tests revealed significant improvements on both measures (see \autoref{fig: ptfs}). PT increased from a pre-training mean of $M = 3.36$ ($SD = 0.65$) to $M = 3.70$ ($SD = 0.51$), $t(21) = 2.98$, $p = .007$, Cohen’s $d = 0.64$ (medium-to-large). FS increased from $M = 2.76$ ($SD = 0.64$) to $M = 3.22$ ($SD = 0.92$), $t(21) = 4.04$, $p < .001$, $d = 0.86$ (large). These results indicate that participants demonstrated stronger perspective-taking and imaginative role engagement after training. In psychological research on human behavior, effect sizes typically range from small to medium ($d = 0.2$–$0.5$)~\cite{schafer2019meaningfulness}, our observed effects ($d = 0.64$–$0.86$) exceed these benchmarks, suggesting that the \textit{CoEmpaTeam} training enhanced cognitive empathy.

\begin{figure}[ht]
  \centering
  \begin{subfigure}[t]{0.45\linewidth}
    \centering
    \includegraphics[width=\linewidth]{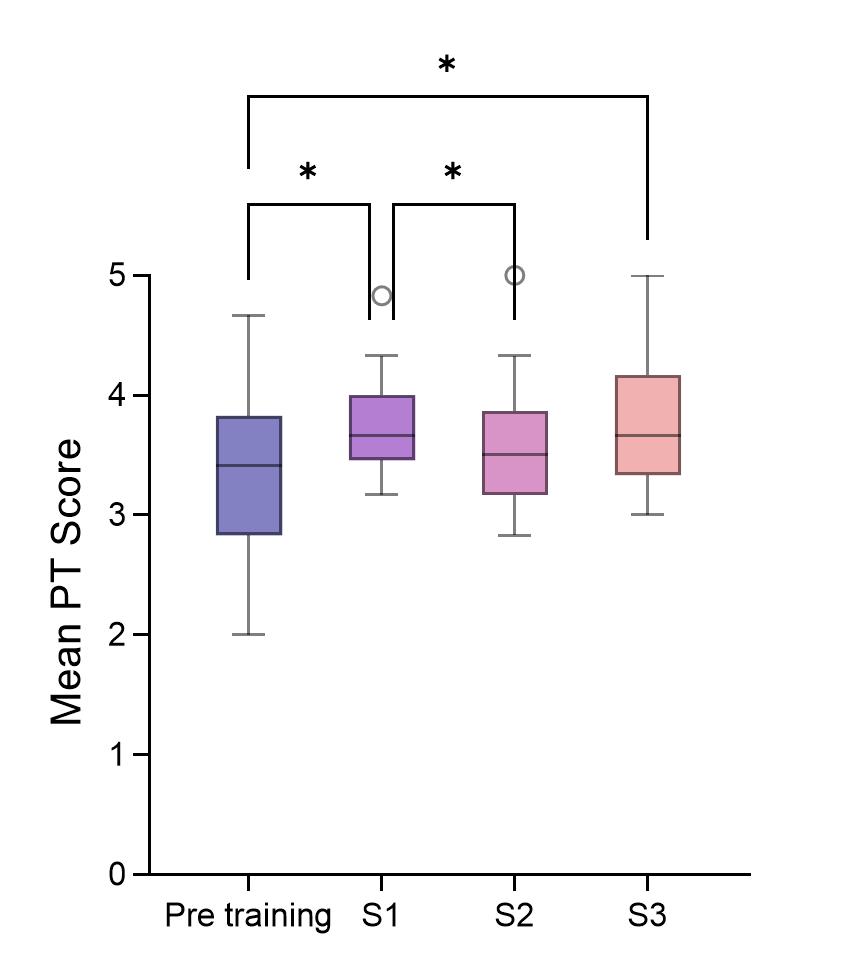}
    \caption{PT}
  \end{subfigure}
  \begin{subfigure}[t]{0.45\linewidth}
    \centering
    \includegraphics[width=\linewidth]{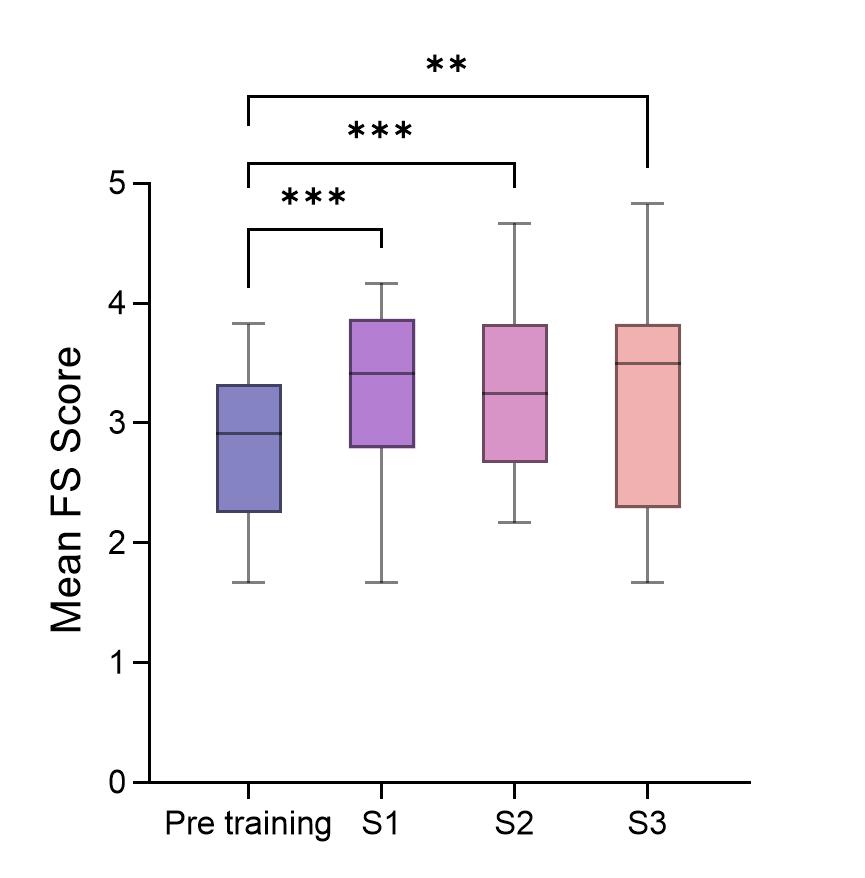}
    \caption{FS}
  \end{subfigure}
  \caption{
    Session-wise IRI scores. PT and FS were elevated relative to pre-training; PT showed session-to-session fluctuations, whereas FS showed no significant differences among the three training sessions.}
  
  \Description{
  Boxplots of Interpersonal Reactivity Index scores across the pre-training baseline and three training sessions. Panel (a) shows PT, which increased from pre-training to Session~1, dipped at Session~2, followed by a non-significant increase at Session~3. Panel (b) shows Fantasy, which was significantly higher than pre-training at all three sessions, with no significant differences among the training sessions.}
  \label{fig: ptfsanova}
\end{figure}

To examine the trajectory of empathy development across sessions, we conducted repeated-measures ANOVAs on PT and FS (see \autoref{fig: ptfsanova}). 
For PT, there was a significant main effect of session, $F(2.11, 44.25) = 7.40$, $p = .0014$, partial $\eta^2 = .26$. Post-hoc comparisons showed that PT increased significantly from pre-training to Session~1, decreased significantly from Session~1 to Session~2, and showed a non-significant increase from Session~2 to Session~3. Despite these fluctuations, PT scores in all training sessions remained higher than at pre-training.
FS also showed a significant main effect of session, $F(2.66, 55.84) = 9.51$, $p < .001$, partial $\eta^2 = .31$. Post-hoc comparisons indicated that FS was significantly higher in all three training sessions compared to pre-training, with no significant differences among Sessions~1, 2, and 3.

Overall, these findings indicate that cognitive empathy improved significantly from pre- to post-training. 
Session-level analyses further reveal a pattern of early gains in both PT and FS, followed by session-to-session fluctuations in PT, while FS remained consistently elevated across all training sessions relative to pre-training.

\paragraph{System Experience: Presence, Embodiment, and Avatar Perception (PQ, VEQ, UVI)}

To assess participants’ experience in VR and their perception of the avatars, we analyzed three measures.
Presence (PQ) scores remained at a moderate level across the three sessions (S1: $M = 3.06$, $SD = 0.42$; S2: $M = 3.20$, $SD = 0.56$; S3: $M = 3.21$, $SD = 0.55$), showing a slight upward trend but no significant differences ($p > .05$) (see \autoref{fig: systemexperience}a). This indicates a stable and sufficient sense of “being there” in VR, maintained as participants became familiar with the system.
Similarly, embodiment scores on the VEQ ownership subscale remained moderate across sessions (S1: $M = 2.46$, $SD = 0.54$; S2: $M = 2.54$, $SD = 0.49$; S3: $M = 2.56$, $SD = 0.57$), with no significant differences ($p > .05$) (see \autoref{fig: systemexperience}b). This limited ownership likely reflects the restricted mapping of movements to the avatar, which included arm control and natural head tracking but not full-body animation. It may also relate to the system’s emphasis on social rather than physical engagement. Nevertheless, the stable ratings suggest that participants could still relate to the avatars in socially meaningful ways. Embodiment may be enhanced in future iterations of \textit{CoEmpaTeam} by incorporating richer body tracking, gesture recognition, or more expressive nonverbal cues.
By contrast, UVI ratings showed no evidence of the classic uncanny valley dip. Instead, comfort increased monotonically with perceived humanness (see \autoref{fig: systemexperience}c). This pattern was consistent across sessions, indicating that participants adapted positively to the avatars’ appearance and experienced them as socially acceptable within the training context.

\begin{figure*}[t]
\label{fig: systemexperience}
  \centering
  \begin{subfigure}[t]{0.28\linewidth}
    \centering
    \includegraphics[width=\linewidth]{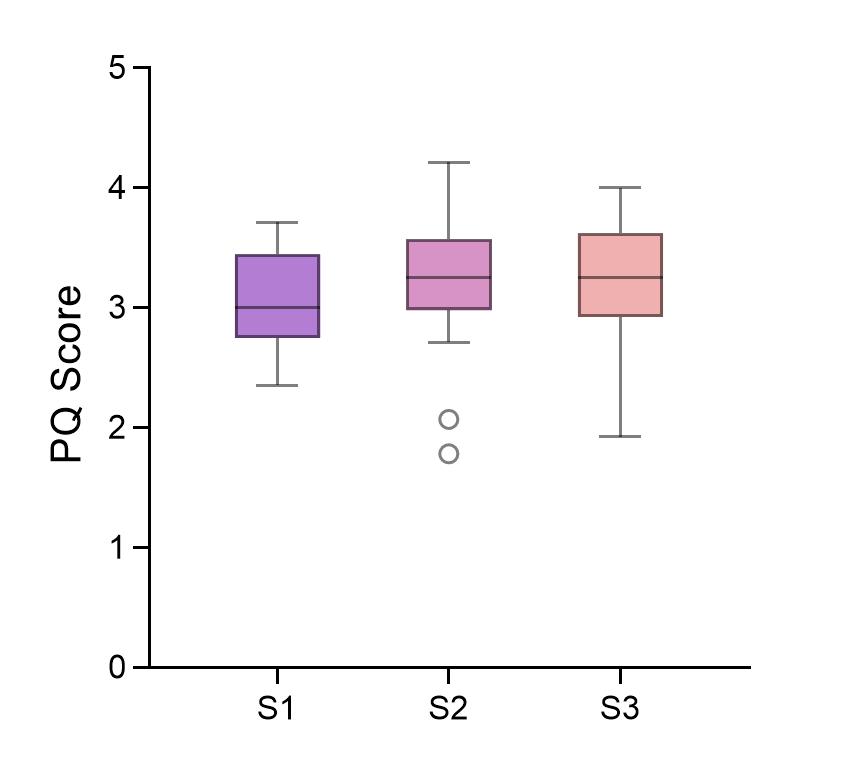}
    \caption{Presence (PQ)}
  \end{subfigure}
  \begin{subfigure}[t]{0.28\linewidth}
    \centering
    \includegraphics[width=\linewidth]{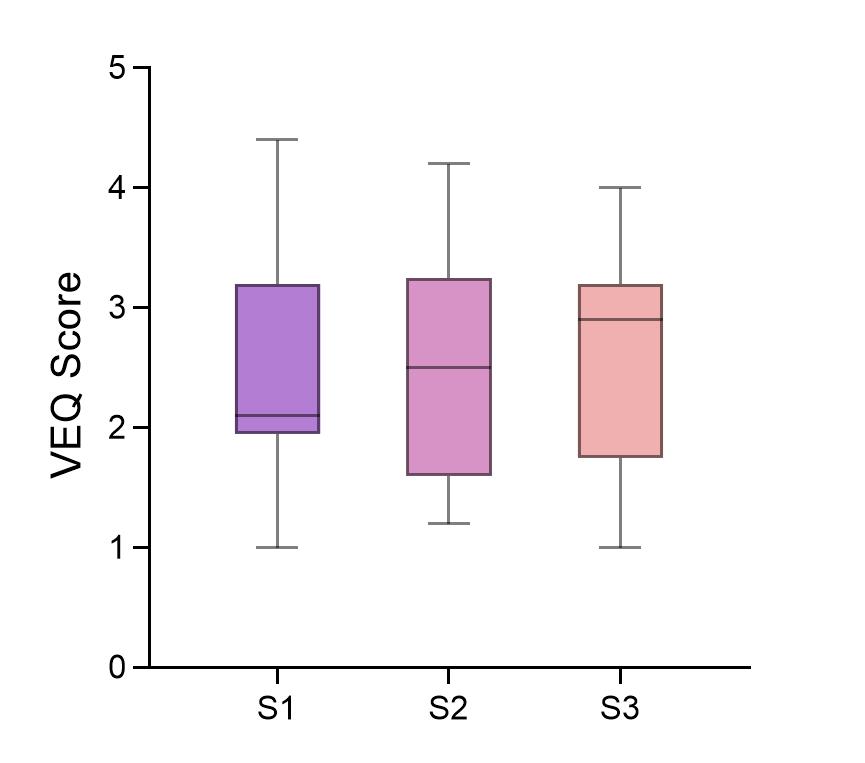}
    \caption{VEQ–Ownership Subscale}
  \end{subfigure}
  \begin{subfigure}[t]{0.4\linewidth}
    \centering
    \includegraphics[width=\linewidth]{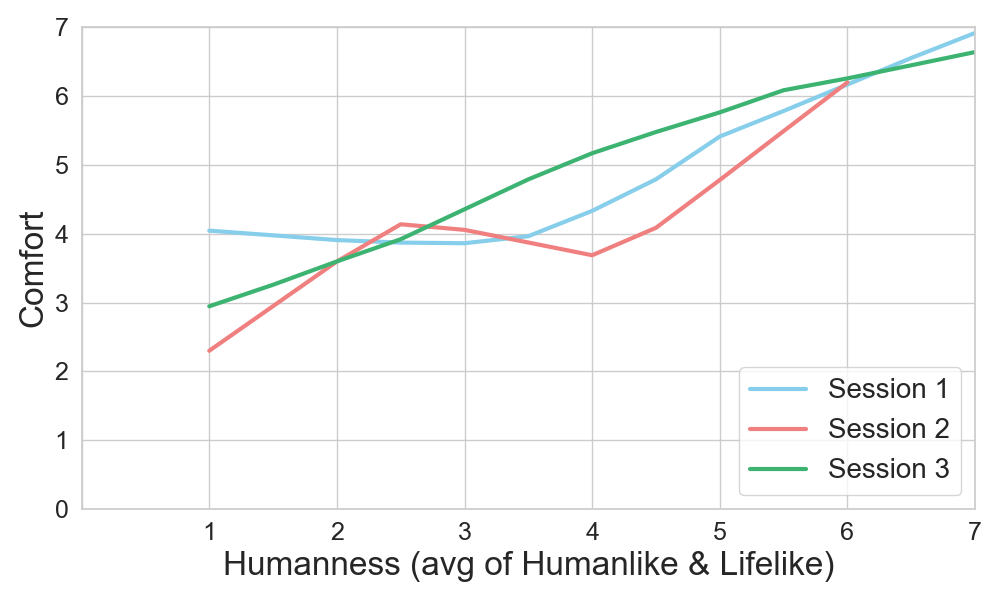}
    \caption{Uncanny Valley Index (UVI)}
  \end{subfigure}
  
  \caption{System experience measures across sessions. (a) Presence remained moderate and stable, (b) Embodiment was consistently moderate, and (c) Comfort increased with humanness without an uncanny valley dip.}
  \Description{System experience measures across three sessions. Panel (a) shows Presence scores with stable moderate values. Panel (b) shows VEQ Ownership scores indicating consistently moderate embodiment. Panel (c) is a line plot of Uncanny Valley Index, where comfort increased steadily with humanness across sessions without an uncanny valley dip.}
  \label{fig: systemexperience}
\end{figure*}

\paragraph{Perceived Workload (NASA-TLX)}

\begin{figure}[ht]
  \centering
  \includegraphics[width=0.75\linewidth]{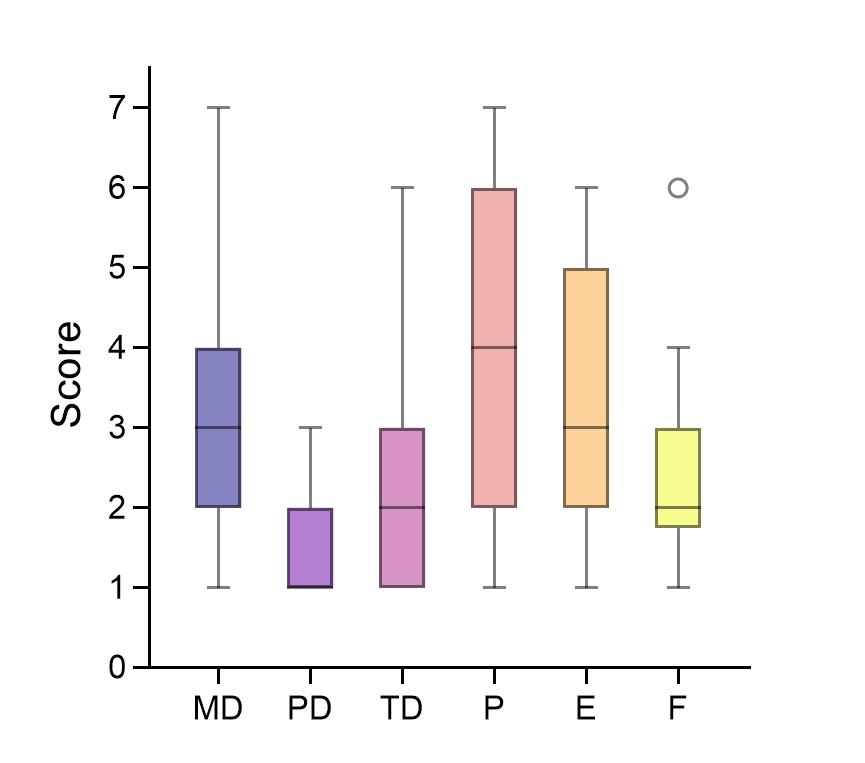}
  \caption{NASA Task Load Index (NASA-TLX) scores across the six subscales: MD, PD, TD, P, E, and F. Ratings indicate moderate mental demand and effort, low physical and temporal demand, high performance, and low frustration.}
  \Description{Boxplots of NASA Task Load Index subscales. Mental demand and effort were moderate, physical and temporal demand were low, performance was rated high, and frustration remained low.}
  \label{fig: nasatlx}
\end{figure}

The NASA-TLX comprises six subscales: Mental Demand (MD), Physical Demand (PD), Temporal Demand (TD), Performance (P), Effort (E), and Frustration (F). Participants rated each subscale on a 7-point Likert scale. For analysis, the Performance scale was reverse-coded so that higher values indicate better performance, while for the other subscales lower values indicate lower workload. Mean scores indicated moderate mental demand ($M = 3.04$, $SD = 1.63$), low physical ($M = 1.50$, $SD = 0.72$) and temporal demand ($M = 2.46$, $SD = 1.56$), high perceived performance ($M = 4.33$, $SD = 2.01$), moderate effort ($M = 3.33$, $SD = 1.43$), and low frustration ($M = 2.21$, $SD = 1.14$) (see \autoref{fig: nasatlx}). Overall, participants were cognitively engaged in the task while experiencing little physical or temporal strain. They perceived their performance as relatively high and reported low levels of frustration, suggesting that the training was demanding enough to elicit involvement without imposing excessive workload.

\subsubsection{Qualitative Findings from Interviews}
\leavevmode\\
Participants completed the training individually. For logistical reasons, two participants were scheduled in the same time slot but worked in separate rooms, while interviews were conducted one-on-one by a single researcher. As a result, only one participant per slot could be interviewed immediately after the session. To obtain a balanced and representative subset, we selected interviewees across time slots while considering gender and cultural diversity. This resulted in 11 completed interviews (P01–P11; $M = 24.27$, $SD = 3.74$; 6 male, 5 female), with cultural backgrounds spanning Western Europe (4), Eastern Europe (1), East Asia (2), South Asia (3), and the Middle East (1). Each interview lasted approximately 20 minutes.
The analysis revealed four recurring themes: Clear frameworks facilitated reflective and enjoyable experiences, design features enabled natural and deeper role adoption, role-switching fostered perspective expansion, and avatars and environments supported empathic engagement. We elaborate on each theme below with illustrative quotes.

\paragraph{\textbf{Clear Frameworks Facilitated Reflective and Enjoyable Experiences.}}   

Participants consistently characterized \textit{CoEmpaTeam} as both enjoyable and reflective. As P01 noted, \textit{“I really enjoyed it, it was a fun experience.”} They highlighted that the system’s clear framework for discussion made the training feel organized rather than arbitrary, which helped them stay attentive and engaged. As P07 noted, \textit{“[The setup] made me stop and think about what the character would do, rather than just answering as myself.”} Similarly, P05 explained that the experience \textit{“was very easy, and as the sessions progressed, I think it was easier for me to adopt [the role].”}
Several participants further appreciated that the task provided a clear framework for dialogue, which they felt supported constructive and thoughtful exchanges. While we ensured that all participants received introductory training to become comfortable with the VR environment, some still felt that their limited familiarity with VR constrained the meaningfulness of the experience. To support novice users, participants suggested adding more intuitive interface cues (P07) and short introductory videos describing the roles to help users settle into the role (P05).
Overall, participants valued the enjoyable and reflective nature of the training, as the clear framework for discussion helped them stay attentive while considering different perspectives. They also suggested improvements to ease onboarding and further enhance the experience in future iterations.

\paragraph{\textbf{Design Features Enabled Natural and Deeper Role Adoption.} }

Participants generally indicated that they were able to step into the assigned roles naturally. For example, P05 noted it was \textit{“like playing a theater role.”} This made role embodiment feel accessible and engaging, allowing them to move beyond their habitual ways of speaking or responding.

Role cards and hidden motives were repeatedly highlighted as essential tools for role adoption. They provided clear background information that anchored participants in the perspective of the role. As P01 explained, \textit{“I was able to do that very naturally because the role cards were very clear.”} They further noted that such traits often resembled people they had encountered in real life, which made the role easier to inhabit. Similarly, P07 added that \textit{“it was pretty easy to follow along because we had the whole [role cards].”}
Hidden motives, in particular, encouraged deeper reflection on how to enact a role. P11 noted, \textit{“when I realized the character had a hidden reason, I thought harder about how to behave.”} Collectively, these design features supported participants in moving beyond surface-level enactment toward more authentic role immersion.

\paragraph{\textbf{Role-Switching Fostered Perspective Expansion.} }
\label{sec: qualitativeroleswitching}

Participants experienced role-switching as both engaging and enlightening. Some likened the activity to a game that nonetheless prompted them to consider others’ perspectives (P01), while others compared it to writing a novel that required imagining characters’ motivations beyond their own (P11). These metaphors illustrate how role-switching combined playful engagement with deliberate perspective-taking, combining enjoyment with cognitive work.

Role-switching also heightened sensitivity to contexts and interpersonal differences. P04 explained that inhabiting different roles forced him to \textit{“think about what others would do in this situation,”} which cultivated greater contextual awareness. Similarly, P09 reflected that \textit{“people think in completely different ways than I do, and I need to pay attention to that,”} noting that fictional roles even reminded him of real-life acquaintances. Such realizations prompted participants to adapt their own preferences or reconsider habitual responses. For instance, P01 recalled adjusting his choices as Benji to accommodate Alice’s early waking habits. Embodying contrasting perspectives side by side highlighted the diversity of thought and behavior, pushing participants to practice perspective-taking beyond what they typically attempted in daily life.

These insights were reinforced through repeated exposure across sessions. P04 noted that sequentially playing multiple roles helped him better grasp the dynamics between roles: \textit{“By switching the roles, I could understand each of them more, when I played Alice, I understood Benji differently, and when I played Caden, I could see both Alice and Benji in new ways.”} Similarly, P07 explained that encountering avatars first as counterparts and later embodying them himself highlighted the contrast between perspectives. He reflected, \textit{“It was interesting to see how they handled it versus how I handled it,”} noting that this shift deepened his perspective-taking across sessions. Together, these accounts suggest that role-switching fostered perspective growth within sessions and gradually consolidated empathic skills through iterative practice over time.

\paragraph{\textbf{Avatars and Environments Supported Empathic Engagement.} }

Avatars were frequently highlighted as important for making interactions feel natural and immersive. Participants described them as \textit{“like real people”} (P04, P05), noting that their gestures and movements enhanced believability. P06 reflected that although the avatars resembled \textit{“in-game characters,”} their actions nonetheless felt \textit{“really natural.”} Voices and facial animations further reinforced immersion; as P03 noted, the voices were \textit{“pleasant,”} while expressions \textit{“gave weight to the dialogue.”} These features provided the believability cues that supported empathic engagement and sustained participants’ connection to their roles throughout the training.

The surrounding virtual environment also played a critical role in fostering presence and empathy. Several participants emphasized how being \textit{“surrounded by the environment”} (P07) made them take the task more seriously, as if they had stepped into the characters’ shared world. P01 appreciated that the setting was \textit{“very clear”} and facilitated interaction, while P04 described that seeing other roommates in VR \textit{“made me more engaged into the conversation, like as if I was in the real world.”} Such accounts suggest that the environment provided a meaningful stage for role-play, strengthening both immersion and empathic resonance.

At the same time, participants remained sensitive to the limitations of avatar nonverbal cues. P10 explained, \textit{“in real life I can read from someone’s face what they think […] but here it was always the same face,”} and P03 noted that this limitation made him rely more on verbal cues. These reflections indicate that while avatars conveyed a range of expressions, their lack of nuance limited participants’ ability to read emotions and sustain empathic engagement. This highlights the need for more refined nonverbal cues in future empathy-oriented systems to better support subtle social signaling.

\subsubsection{Qualitative Findings from Diary}
\leavevmode\\
Sixteen participants (D01–D16; $M=23.62$, $SD=3.03$; 11 male, 5 female) ultimately completed the entries. The thematic analysis of these diaries revealed three recurring patterns that illustrate how cognitive empathy training was reflected in daily life, through noticing differences in perspectives, adapting responses, and sustaining cognitive empathic practices beyond the VR sessions.

\paragraph{\textbf{Recognizing Perspective Differences in Everyday Interactions.} } 

Participants reported becoming more attentive to differences in perspectives, needs, or priorities during everyday interactions. This awareness arose in social situations, ranging from casual conversations with friends to shared responsibilities or collaborative tasks, where participants reflected on how their own actions might be experienced from different standpoints. As D07 explained: \textit{“I try to understand how others feel when they are criticized. Because I once criticized others for not completing their tasks in a project.”} Beyond immediate interactions, some participants also extended this awareness to online contexts. D09 described reflecting on polarized audience reactions to a TV series episode: \textit{“While I personally enjoyed the latest episode, many people had strong negative reactions […] I read more carefully about those negative comments that are genuine discussions with their thoughts and reasons, not just emotional outbursts. I became interested in learning others' perspectives, even if they may conflict with my own.”} These accounts suggest that participants increasingly noticed and considered diverse perspectives across both everyday and mediated interactions.

\paragraph{\textbf{Adjusting Communication and Reactions with Empathic Strategies.}} 

Beyond noticing perspective differences, participants described actively adapting their communication and reactions to foster more constructive interactions. Several noted becoming calmer and more patient in daily exchanges. As D02 reflected: \textit{“In general I have become calmer and more understanding of other people’s viewpoints. ”} Similarly, D03 explained how empathic reflection helped resolve a disagreement with their partner: \textit{“I am constantly reminded of the sessions and how important it is to check each other’s perspectives before coming to a conclusion […] Just today my girlfriend and I had an argument about our schedules, but we talked it through, and I think my patience has increased.”} Other participants echoed this emphasis on constructive responses, highlighting greater willingness to listen and adjust how they communicated. D13 described applying these strategies in routine contexts: \textit{“When my friends and I were discussing the dinner menu, I listened to their ideas first before questioning them. This helped me consider different perspectives and realize that my own view might be wrong.”} D09 explained how they sought to balance perspectives: \textit{“I try to get more information before making the judgement and think of whether there is room for compromising or even a win–win situation.”}

These accounts illustrate how participants drew on cognitive empathy to guide their communication and responses, applying it to both personal disagreements and mundane situations, thereby fostering constructive and cooperative interactions in daily life.

\paragraph{\textbf{Sustaining Cognitive Empathy Across Time and Contexts.}} 

Participants reported that the practices introduced in VR carried over into a wide range of daily situations, from study sessions and household negotiations to personal disagreements. For example, D04 explained: \textit{“It helped me better to understand how different people may be, and if we need to decide something together we need to take the difference into consideration for better communication. I try to use this knowledge when I am arguing with someone.”} Likewise, D06 emphasized applying perspective-taking to regulate emotional reactions: \textit{“When I am upset about someone’s behavior, I try to understand other people’s perspectives […] it helps in controlling impulsive reactions and showing more patience.”} Beyond specific incidents, participants also highlighted the longer-term relevance of these strategies. D08 noted: \textit{“Because they are easy to remember and they work.”} D09 further linked them to relationship quality: \textit{“Because it makes the relationship between people more harmonious.”} These accounts suggest that the effects of VR training extended beyond the sessions themselves, supporting the sustained use of empathic practices across diverse contexts and over time.

At the same time, a few participants indicated that they did not perceive major changes, as they already considered themselves strong in perspective-taking. These accounts suggest that while the training fostered sustained and transferable applications of cognitive empathy for many, its impact varied depending on participants’ prior dispositions.

\subsection{Summary}

Overall, Study 2 demonstrated that \textit{CoEmpaTeam} effectively enhanced participants’ cognitive empathy. Quantitative results showed significant early gains in both perspective-taking and fantasy. While fantasy remained consistently elevated across all training sessions, perspective-taking showed early improvement followed by session-to-session fluctuations, yet remained higher than at pre-training throughout the study. Complementary qualitative findings revealed that participants not only engaged in perspective-taking during VR interactions but also extended these practices into everyday contexts. These results suggest both immediate improvements and meaningful transfer of cognitive empathy beyond the VR setting. In the following section, we discuss the mechanisms underlying these outcomes and their implications for the design of cognitive empathy-focused training systems.

\section{Discussion}

In this section, we interpret our findings by discussing the mechanisms that may explain these effects, the design implications for cognitive empathy-focused training systems, and the broader relevance of our work for HCI.

\subsection{How Role-Switching with LLM-Driven Avatars Cultivates Cognitive Empathy}

Our findings suggest that role-switching in VR fostered cognitive empathy by engaging participants in both enactment and observation. When embodying a role such as Alice, participants not only interacted with other avatars but also observed their behaviors from a distinct standpoint. In subsequent sessions, switching into these roles required participants to reinterpret the same situation from new perspectives, for example, playing Benji after previously observing him as Alice. This process of alternating between self-as-character and observer-of-others reflects principles from social cognitive theory~\cite{bandura1986social}, where enactive mastery and vicarious observation jointly contribute to skill development, creating structured opportunities to practice perspective-taking.

Qualitative findings offer process-level indications of this mechanism. Participants described adopting each character’s perspective, comparing these viewpoints with those they had embodied in earlier roles, and updating their understanding of the situation after switching roles. These perspective-shifting and interpretive-updating processes align with the core cognitive operations of cognitive empathy, indicating that the role-switching mechanism functioned as intended within the VR system powered by LLM-driven avatars. The stability of role identities established in Study~1 further reinforced these effects: the avatars consistently expressed coherent personalities, enabling participants to ground their reasoning in stable character traits rather than in arbitrary variation.

Further, beyond the immediate engagement, repeated switching roles across the three sessions enabled participants to revisit familiar situations from multiple angles, deepening their interpretive understanding over time.
As P04 explained: \textit{“By switching the roles, I could understand each of them more, when I played Alice, I understood Benji differently, and when I played Caden, I could see both Alice and Benji in new ways.”} This cycle of enactment, observation, and reflection aligns with experiential learning theory~\cite{kolb2014experiential}, where iterative exposure transforms momentary exercises into sustained skill development.

Alongside these qualitative developments, the quantitative PT trajectory showed session-to-session fluctuations, with a temporary dip at Session 2. Research on skill acquisition indicates that rapid early gains are often followed by short-term performance variability as learners transition from initial exposure into deeper processing~\cite{schmidt1992new, kruger1999unskilled}. Experiential learning theory likewise suggests that new learning situations can produce strong early engagement~\cite{kolb2014experiential}. This aligns with the marked increase observed in Session 1, which may have been amplified by participants’ limited prior VR experience. As novelty diminished and participants engaged more analytically with the task, ratings dipped temporarily rather than showing a sustained decline. Such mid-trajectory fluctuations are consistent with consolidation processes in skill development~\cite{ericsson1996expert} and with calibration effects in which learners develop more accurate self-assessments~\cite{kruger1999unskilled}. Despite these fluctuations, PT scores remained above pre-training levels, indicating that the overall improvements in cognitive empathy were retained beyond novelty-driven effects.

In addition to cognitive gains, we observed that participants occasionally reported emotional reactions during the sessions. This is consistent with research showing that perspective-taking can indirectly evoke affective response~\cite{healey2018cognitive}, as reflecting on others’ situations may elicit emotional reactions. However, the system was intentionally designed to cultivate cognitive rather than affective empathy. Following Davis’s formulation~\cite{davis1983measuring}, cognitive empathy concerns understanding another person’s thoughts and motivations without requiring shared emotional states. 
To maintain this focus, the scenario structure, LLM-driven avatars, and role-switching in \textit{CoEmpaTeam} were crafted to emphasize perspective-taking and reflective reasoning, processes central to cognitive empathy.

Conceptually, our findings extend prior work that emphasized narrative immersion and emotion as primary drivers of empathy~\cite{rifat2024cohabitant, kors2020curious}. Participants’ interactions with avatars resembled parasocial relationships~\cite{horton1956mass}, in which people perceive mediated characters as socially real. However, unlike traditional one-sided parasocial interaction, the LLM-driven avatars adapted their dialogue to participants’ input, creating reciprocal exchanges that made the interaction feel socially responsive. This shift illustrates how VR systems can move beyond offering momentary empathic experiences to functioning as environments for cultivating cognitive empathy as a transferable skill.

Building on this distinction between passive immersion and interactive engagement, it is also important to consider how VR compares to traditional in-person role-play. While similar role-switching could, in principle, be conducted with actors or with multiple participants enacting the respective roles, our VR-based approach provides scalability, consistent scenario delivery, and adaptive interactivity. Within the same scenario, LLM-driven avatars flexibly adapt their dialogue while maintaining stable character identities, eliminating the need for multiple actors or participants. These affordances position \textit{CoEmpaTeam} as a complement to traditional in-person training, particularly valuable when participant numbers or resources are limited. In this way, VR functions not as a replacement for human role play but as a scalable and controlled medium for structured empathy practice and transferable skill development.

\subsection{Extending Cognitive Empathy Practices Beyond VR}

Our findings show that cognitive empathy practices cultivated through \textit{CoEmpaTeam} extended into participants’ everyday interactions. Like prior work~\cite{rifat2024cohabitant, winters2021can}, we collected immediate self-reports after each session, but we complemented this with a week-long diary that captured participants’ reflections in situ. 
Because our evaluation was situated within a single co-living scenario, the diary served as an important complement to in-session measures by revealing how participants applied perspective-taking in daily negotiations and social encounters, and how they reflected on these experiences.
This resonates with reflective learning theories~\cite{boyd1983reflective}, which highlight reflection as a bridge from situated experience to transferable practice.
While such accounts cannot replace multi-scenario evaluations, they illustrate how cognitive empathy practices developed in VR may be enacted beyond the immediate training context.

Overall, these results demonstrate that empathy training outcomes were not confined to controlled sessions but also manifested in real-life contexts, thereby enhancing the ecological validity of our findings. However, it is important to distinguish short-term improvements from long-term learning effects. Prior work shows that structured perspective-taking activities can yield meaningful short-term gains in cognitive empathy~\cite{lam2011empathy,riess2012empathy}, whereas sustaining such gains typically requires continued practice, contextual reinforcement, or ongoing reflective engagement~\cite{riess2012empathy}. This distinction highlights a well-recognized challenge in empathy research: short-term gains often stem from situationally activated perspective-taking, whereas long-term change requires the gradual consolidation of interpretive habits across varied social contexts. Our evaluation provides preliminary evidence of short-term transfer into everyday interactions, but it does not address long-term durability or broader generalization beyond the study period. Future work should investigate how systems like \textit{CoEmpaTeam} can be integrated into longer-term training programmes or everyday digital platforms to support sustained development over time.

\subsection{Design Implications for Cognitive Empathy Training Systems}

Our findings yield several design insights for the development of systems that aim to foster cognitive empathy.

\subsubsection{Structured Cognitive Tasks}

While immersive storytelling can make empathy training engaging~\cite{xu2024istraypaws}, our results show that cognitive empathy benefits most when participants are required to reason explicitly about different viewpoints. In our case, negotiating household rules prompted them to articulate and reconcile divergent perspectives, turning immersion into active reflection. To fully realize the potential of VR for cultivating cognitive empathy, designs should therefore move beyond immersion and narrative alone, embedding explicit cognitive interventions that make perspective-taking and critical reflection central to the experience~\cite{kukshinov2025seeing}.

\subsubsection{Distinct and Consistent Role Design}

The credibility of the training relied on avatars whose personalities and behaviors were both stable and recognizable. When participants could clearly identify a role’s traits and motivations, they found it easier to distinguish between perspectives and adjust their own responses accordingly. Empathy training systems should thus prioritize character design that conveys distinct and consistent identities, expressed through personality traits, behavioral cues, and dialogue patterns, to scaffold reliable and meaningful perspective-taking.

\subsubsection{Relatable Everyday Contexts}

Situating the training in familiar interpersonal scenarios, such as household negotiations, made it easier for participants to apply perspective-taking strategies beyond VR. Because these scenarios mirrored the kinds of interactions they routinely encountered with roommates, friends, or classmates, participants could more readily see how empathic reflection was relevant to their own lives. This finding aligns with our initial assumption that embedding everyday conflicts into the training provides a meaningful basis for eliciting perspective-taking and, in turn, fostering the development of cognitive empathy. Empathy training systems should therefore ground practice in relatable social contexts that connect directly to users’ lived experiences, increasing the likelihood that strategies rehearsed in VR transfer into everyday interactions.

Taken together, these implications suggest that VR-based cognitive empathy training benefits from structured perspective-taking tasks, persona-consistent roles, and relatable everyday contexts. In \textit{CoEmpaTeam}, repeated role switching across three avatars supports enactment, observation, and reflection across sessions.

\subsection{Scalability Considerations in Avatar Persona Design}

A related systems-level consideration concerns scalability, especially in designing and validating persona-consistent avatars. In our implementation, creating coherent role profiles and ensuring behavioral consistency involved manual curation, which constrains scaling to larger deployments or rapid adaptation to new domains. 
For verbal behavior, recent advances in LLM-based Big Five persona shaping~\cite{li2025big5} demonstrate that linguistic style can be systematically aligned with target personality traits, offering initial steps toward reducing reliance on manually crafted personas. 
For nonverbal behavior, research in social signal processing has identified foundational links between expressive cues and social perceptions~\cite{VINCIARELLI20091743}. However, existing taxonomies are not yet sufficiently formalised or computationally actionable to support fully automated persona-consistent avatar generation. Developing such taxonomies and exploring how they might be combined with persona-shaped language models represents a promising direction for improving the scalability of multi-role systems.  
The modular architecture of \textit{CoEmpaTeam}, which separates role descriptions, task structure, and dialogue generation, provides a practical foundation for incorporating these advances. As computational models of verbal and nonverbal persona expression mature, new roles or scenarios could be integrated without redesigning the overall system, reducing the dependence on human validation and enabling broader applicability.

\subsection{Broader Implications for HCI}

Although our study was grounded in co-living negotiations, the approach may inform a wider range of HCI contexts. At the system level, \textit{CoEmpaTeam} suggests that LLM-driven avatars could function not only as conversational agents but also as structured partners that support perspective-taking. While we instantiated this approach in a household scenario, similar designs may be adaptable to domains such as teamwork training, intercultural communication, or healthcare education, where understanding multiple perspectives is essential. The same framework could also extend beyond cognitive empathy to other soft skills, such as conflict resolution or collaborative decision-making, by tailoring role content and task dynamics.

Our study highlights the value of pairing controlled VR training with longitudinal methods as one way to explore whether training effects extend beyond the lab. By combining in-session measures with a week-long diary, we were able to capture both immediate experiences and short-term transfer, showing how established qualitative methods can complement controlled interventions. 

Finally, our study contributes to conversations about the role of empathy in HCI. Prior work has treated empathy as an experience, for example, immersing users in a designed scenario to momentarily adopt another’s perspective~\cite{kishore2019virtual}. In contrast, our findings suggest that interactive systems can serve as training environments that cultivate empathy as a transferable skill. This perspective shifts the focus from designing for one-time empathic experiences to designing for empathic capacities, equipping users with abilities that endure and can be applied across contexts and time.

\subsection{Ethical Considerations}

LLM-driven dialogue systems involve widely recognized ethical risks, particularly the potential to generate biased or stereotypical responses~\cite{fang2025social, gallegos2024bias}. In the context of cognitive empathy training, such risks are critical to address, as they can undermine the very goal of fostering perspective-taking. In \textit{CoEmpaTeam}, avatars were intentionally crafted with distinct personalities to represent contrasting viewpoints, making behavioral differences part of the design. However, LLM-generated dialogue could still risk introducing unintended stereotypes—for instance, associating assertiveness or passivity with gender or cultural traits—that go beyond these intended characteristics. Although our study did not collect personal data and thus posed minimal privacy risks, inappropriate or exclusionary outputs could nonetheless cause discomfort. Future systems should therefore incorporate safeguards such as auditing and calibrating dialogue, as well as mechanisms for user feedback and disengagement, helping to ensure that designed role differences foster empathy without inadvertently reinforcing stereotypes.

\section{Limitations and Future Work}

While our findings provide evidence for the effectiveness and transfer of cognitive empathy training in VR, several limitations warrant consideration. 

First, our sample was relatively homogeneous, consisting mainly of young university students (average age 24.39). This may limit the generalizability of the findings to populations, particularly other age groups or those less familiar with immersive technologies. 
Beyond age homogeneity, cultural differences were not examined. 
Although our sample included participants from multiple cultural groups, we did not analyze how cultural norms might shape role interpretation, perspective-taking strategies, or interactions with LLM-driven avatars. 
Given that cultural factors influence empathic reasoning and communication styles~\cite{markus2014culture}, future work could investigate how users from different cultural contexts engage with role-switching and whether culturally grounded adaptations of role profiles or dialogue patterns enhance the training experience. 
This limitation also points toward opportunities for system adaptivity. 
Future systems could leverage LLMs to dynamically adjust avatar characteristics to create perspective-taking challenges tailored to each user. 
Such adaptations could align with a user’s cultural background or deliberately introduce contrasting or culturally distinct viewpoints, depending on the pedagogical goal, enabling more culturally and personally adaptive empathy-training experiences.

In addition, the study did not include a non–role-switching baseline. This aligns with our evaluation focus: \textit{CoEmpaTeam} integrates role-switching, personality-driven LLM avatars, and embodied VR interaction into a single, cohesive training experience, and the present study examines the effectiveness of the system as a whole rather than isolating individual components. To support further research, we open-source the system to facilitate reproducibility and enable future work to develop comparative variants or baseline-controlled implementations.

Another limitation concerns technical constraints. Although advances in LLMs and speech recognition enabled fluid avatar interactions, occasional latency still disrupted conversational flow. Prior work suggests that delays longer than about four seconds can negatively impact user experience~\cite{maslych2025mitigating}. Future systems could adopt established strategies to mitigate this issue, such as incorporating hesitation gestures (e.g., touching the chin) or verbal fillers (e.g., short reflective phrases) to simulate human-like pauses and preserve conversational continuity.

Our evaluation approach also presents limitations, as it relied primarily on self-reports, which capture subjective experiences but may not fully reflect the complexity of empathic processes. Complementary biosignals, such as heart rate variability, respiration, or galvanic skin response, could provide insights into how cognitive empathy unfolds during and after training~\cite{lee2022understanding, parra2022virtual}. Beyond serving as measurement tools, biosignals also hold potential as expressive media in social VR. Because avatars' nonverbal cues, such as facial expressions, are often less nuanced than in real life, integrating biosignal-based feedback (e.g., visualizing heart rate or breathing rhythm through avatars) could offer an additional channel for conveying affective and cognitive states, supporting more natural and empathic communication~\cite{lee2022understanding}.
Building on these possibilities, future extensions of \textit{CoEmpaTeam} could explore affective aspects of empathy more directly, for example by incorporating avatars capable of expressing affective cues and by leveraging biosignal-based visualization to enrich emotional expressivity in VR.

Finally, our role design was limited to a co-living scenario with three predefined avatars. While this served as a proof of concept, the framework can be extended to roles representing diverse demographics, cultural backgrounds, or even physical conditions, allowing users to embody perspectives further removed from their own~\cite{zhang2025inclusive}. In future work, supporting customizable role configurations and scenario design could enhance engagement and adaptability, enabling the system to extend beyond empathy training toward the cultivation of broader soft skills, such as conflict resolution, collaborative decision-making, and other transferable competencies. Overall, these directions highlight opportunities for \textit{CoEmpaTeam} to evolve from the current implementation into a flexible platform for fostering transferable skills across diverse domains.

\section{Conclusion}

We introduce \textit{CoEmpaTeam}, a novel VR-based system that leverages LLM-driven avatars with distinct personalities and a structured role-switching mechanism to facilitate cognitive empathy training. In Study 1, we validated through avatar self-assessment and human evaluation that the three designed avatars reliably expressed their intended personalities, providing a credible foundation for subsequent cognitive empathy training. In Study 2, we showed that repeated role-switching improved participants’ perspective-taking and fantasy scores, with the acquired skills transferring into real-world contexts as captured in a diary study. Collectively, these findings highlight the potential of structured role-switching in VR to cultivate cognitive empathy, and we hope they can inform future HCI research on designing interactive systems that foster cognitive empathy skills.

%%
%% The acknowledgments section is defined using the "acks" environment
%% (and NOT an unnumbered section). This ensures the proper
%% identification of the section in the article metadata, and the
%% consistent spelling of the heading.

%%
%% The next two lines define the bibliography style to be used, and
%% the bibliography file.
\bibliographystyle{ACM-Reference-Format}
\bibliography{References}

%%
%% If your work has an appendix, this is the place to put it.

\end{document}